# Topological Bonding and Electronic properties of $Cd_{43}Te_{28}$ semiconductor material with microporous structure

Yixin Li[1], Wei Xiong[1], Lei Li[1], Zhuoming Zhou[1], Chuang Yao[1], Zhongkai Huang[1], and Maolin Bo[1]*

*Corresponding Author: E-mail: bmlwd@yznu.edu.cn (Maolin Bo).

**Abstract:**

CdTe is II-VI semiconductor material with excellent characteristics and has demonstrated promising potential for application in the photovoltaic field. The electronic properties of $Cd_{43}Te_{28}$ with microporous structures have been investigated based on density functional theory. The newly established binding-energy and bond-charge model have been used to convert the value of Hamiltonian into bonding values. We provide a method for describing topological chemical bonds by atomic coordinates and wave phases. We also discuss the dynamic process of the wave function with time and the magic cube matrix. This study provides an innovative method and technology for the accurate analysis of the topological bonding and electronic properties of microporous semiconductor materials.

[1]Key Laboratory of Extraordinary Bond Engineering and Advanced Materials Technology (EBEAM) of Chongqing, Yangtze Normal University, Chongqing 408100, China

## 1. Introduction

Microporous materials are solids that involve regular pores or voids that are less than 2 nm in diameter.[1] Metal-organic frameworks and zeolite are important components of the microporous material category. Most of the microporous materials are composed of connected $TO_4$ tetrahedrons (where T = tetrahedral atoms, such as Al, P, Si, etc.), wherein each oxygen atom is shared between two ring tetrahedrons, thus forming a framework with an O/T ratio of 2 . [2, 3] These tetrahedrons are connected in such a way that regular pores and channels are formed in the material; therefore, a large portion of the material (up to 50% in some cases) is a real "space." Pore systems of 1, 2, or 3-dimension(s) are formed inside the microporous material, and different molecules can be separated according to the geometric shapes of the microporous material. The geometric shape of the microporous material is determined by the size and uniformity of the pores, which is termed "shape selectivity." By controlling the shape selectivity, specific reactant molecules enter the microporous material, and the molecules that can passed into or form in the material are restricted (reactant selectivity) [4] .Similarly, specific products can be removed from materials (product selectivity), and transition states can be formed by specific products in the materials (transition state selectivity) is based on this property. With the discovery and development of new types of materials, the applications, properties, and modifications related to microporous materials have become the focus of scientific research. The increased impetus of the rate of research is due to the excellent performance of microporous materials.

Recently, II-VI semiconductor compounds and alloys that are based on Cd have received extensive attention because of their applications in photoelectric devices[ 5-7].Cadmium telluride (CdTe) is a *p*-type semiconductor with a high absorption rate and wide band gap. Thus, it can be used as an effective material in solar thin film cells, as it exhibits satisfactory efficiency and is cost effective. The working efficiency of the solar thin film cells that comprise CdTe can be greater than 22% [8]. The wide band gap and high absorption rate of CdTe meet the requirements of solar thin film cells. The absorption edge of CdTe is sharp due to the direct band gap [9], and over 90% of the incident light can be absorbed by the material. The maximum photo-current of the CdTe thin film solar cells reaches 30.5 mA/cm$^2$ under an irradiation of 100 mW/cm$^2$ optical power density, and the value of the theoretical maximum efficiency exceeds 27%.[10].Owing to its vast application prospects, the CdTe semiconductor has reached a level comparable with that of single crystal materials that are more complex [11].

Numerous theoretical and experimental studies have been performed on CdTe semiconductor materials [12-14]. Herein, we report the CdTe semiconductor materials that have a microporous structure. The geometric structure of $Cd_{43}Te_{28}$ microporous material is obtained by self-assembly of nano-cluster structure. The geometric structure of different types of $Cd_{43}Te_{28}$ has been optimized using density functional theory (DFT) calculations. Furthermore, the bonding and electronic properties of $Cd_{43}Te_{28}$ semiconductor microporous materials have been studied. The newly established binding-energy and bond-charge (BBC) model has been used to convert the value of Hamiltonian into the values of nonbonding, bonding, and antibonding, and the relationship between the relevant physical and chemical performance parameters of the materials and their chemical bonding was ascertained.[15-17]

The BBC model is quantified chemical bonds by binding energy shift and deformation charge density. For the binding energy shift, we use the central field approximation [18] and Tight-binding (TB) model [19], which can get the relationship between the energy shift and Hamiltonian. For the bond-charge model, we use the second-order term of energy expansion and use deformation charge density to calculate bonding states. For the topological structure, we use the fiber worm principle of quantum mechanics and the geometric phase of Berry.[20] .In theory, by regulating the unique microporous structure and bonding performance of the semiconductor, the photoelectric properties of $Cd_nTe_m$ can be improved markedly.

## 2. Methods

### 2.1 Density functional theory calculations

All the structural relaxation and electronic properties of $Cd_{43}Te_{28}$ microporous materials was investigated with the Cambridge Sequential Total Energy Package (CASTEP) [21], which used DFT with a plane-wave pseudopotential. This analysis was focused on analyzing the atomic structure, energetics, and electronic properties of $Cd_{43}Te_{28}$ microporous materials. We used the local-density approximation (LDA) and Perdew–Burke–Ernzerhof (PBE) to describe the electron exchange and correlation potential [22-24] .The cutoff energy of the plane-wave basis set was 400 eV. The *k*-point grids were 3 × 3× 3, as shown in **Table 1**. In the calculations, all the atoms were completely relaxed such that the energy converged to $10^{-6}$ and the force on each atom converged to <0.01 eV/ Å.

**2.2 BBC model**

The TB model leads to the Hamiltonian formulas [25-28]:

$$\begin{cases} H = -\dfrac{\hbar^2\nabla^2}{2m} + V_a(\vec{r}) + V_c(\vec{r})(1+\Delta_{\mathrm{H}}) \\ V_c'(\vec{r}) = V_c(\vec{r})(1+\Delta_{\mathrm{H}}) = \gamma V_c(\vec{r}) \end{cases} \tag{1}$$

$$\begin{cases} E_v(0) = -\left\langle \phi_v(\vec{r}) \right| -\dfrac{\hbar^2\nabla^2}{2m} + V_a(\vec{r}) \left| \phi_v(\vec{r}) \right\rangle \\ E_v(x) - E_v(0) = -\left\langle \phi_v(\vec{r}) \middle| V_c'(\vec{r}) \middle| \phi_v(\vec{r}) \right\rangle - \sum_j f(k) \left\langle \phi_v(\vec{r}) \middle| V_c'(\vec{r}-\vec{R}_j) \middle| \phi_v(\vec{r}-\vec{R}_j) \right\rangle \\ \qquad = -\left\langle \phi_v(\vec{r}) \middle| V_c'(\vec{r}) \middle| \phi_v(\vec{r}) \right\rangle \left[ 1 + \dfrac{\sum_j f(k) \left\langle \phi_v(\vec{r}) \middle| V_c'(\vec{r}-\vec{R}_j) \middle| \phi_v(\vec{r}-\vec{R}_j) \right\rangle}{\left\langle \phi_v(\vec{r}) \middle| V_c(\vec{r}) \middle| \phi_v(\vec{r}) \right\rangle} \right], \\ \qquad = \gamma\alpha_v \left(1 + \dfrac{\sum_j f(k)\cdot\beta_v}{\alpha_v}\right) \approx \gamma\alpha_v \propto E_x \end{cases} \tag{2}$$

$$\begin{cases} \alpha_v = -\left\langle \phi_v(\vec{r}) \middle| V_c(\vec{r}) \middle| \phi_v(\vec{r}) \right\rangle \\ \beta_v = -\left\langle \phi_v(\vec{r}) \middle| V_c(\vec{r}-\vec{R}_j) \middle| \phi_v(\vec{r}-\vec{R}_j) \right\rangle \end{cases} \tag{3}$$

where $E_v(x)$ is the $v$-th energy level of a crystal and $E_v(0)$ is the $v$-th energy level of an isolated atom. $V_a(\vec{r})$ and $V_c(\vec{r})$ are the potential energies of the atom and crystal, respectively; $\phi_v(\vec{r})$ represents an atomic orbital of the $v$-th energy level; $E_x$ represents the single bond energy. Both $\alpha_v$ and $\beta_v$ in Formula 2 contribute to the energy band width. However, the energy band of the core energy level is determined by $\alpha_v$, because $\beta_v$ is very small in the local energy band of the core energy level, the term $\frac{\sum_j f(k)\cdot\beta_v}{\alpha_v} \ll 1$. The wave function is $\psi_k(\vec{r}) = \frac{1}{\sqrt{N}}\sum_l e^{ik\vec{R}_j}\phi(\vec{r}-\vec{R}_j)$. The form of the periodic factor $f(k)$ is $e^{ik\vec{R}_j}$, where $k$ represents the wave vector. The quantity $\beta$ depends on the overlap between the orbitals centered at two neighboring atoms; $\vec{r}$ represents the electron coordinates and $\vec{R}$ represents the nuclei coordinates. Considering the time-dependent potential function:

Considering the dynamic process of potential function:

$$\begin{cases} V_i(\vec{r},t) = (1+f(t))V_0(\vec{r}) \\ \zeta(f(t)) = F(s) = \int_{-\infty}^{+\infty} f(t)e^{-st}dt \\ f(t) = \zeta^{-1}(F(s)), \hat{S}_t = e^{-iEt/\hbar} = e^{-iwt} \end{cases} \tag{4}$$

$\zeta$ is Laplace transform. $\hat{S}_t$ is an operator. The $s=\kappa+jw$ is a complex number. When $\kappa$ is 0, the function $f(t)$ is Fourier transform. **Table 2** is the Laplace transform formula.

The inner electrons are affected by the potential energy of the crystal, which causes shifts in the core energy levels. By combining the initial- and final-state effects with the TB model, we obtain:

$$\begin{cases} V_c'(\vec{r}) = V_c(\vec{r})(1 + \Delta_H) = \gamma V_c(\vec{r}) \\ V_a(r) = -\frac{1}{4\pi\varepsilon_0}\frac{Z'e^2}{\vec{r}_i}, V_c(\vec{r}) = -\sum_{i,j,R_j\neq 0}\frac{1}{4\pi\varepsilon_0}\frac{Z'e^2}{\left|\vec{r}_i - \vec{R}_j\right|} \end{cases} \tag{5}$$

The core electron will not only be attracted by the nuclear charge, but also be excluded by other electrons. The repulsive effect of electrons will reduce the attractive effect of the nucleus. The effective positive charge of the ion is $Z' = Z - \sigma$, considering the charge shielding effect ( $\sigma$ ), where $Z$ is the nuclear charge. $\sigma = \left|\vec{r}_i - \vec{R}_j\right| / \left|\vec{r}_i - \vec{r}_j\right|$ can be written as a Hamiltonian containing electron interaction terms. However, considering the repulsion of electrons, the effective positive charge number $Z'$ of the ion can be varied, as shown in **Fig. 1**. When the valence electron density decreases, the shielding effect will be weakened and the binding energy of the core electrons will increase. On the contrary, the binding energy will decrease.

$$\Delta E_v(x) = E_v(x) - E_v(0) = -\langle \phi_v(r) | V_c(\vec{r})(1 + \Delta_H) | \phi_v(r) \rangle = -\langle \phi_v(r) | \gamma V_c(\vec{r}) | \phi_v(r) \rangle \tag{6}$$

From Formula 2 and Formula 5, we obtain Formula6:

$$\Delta E_v(B) = E_v(x) - E_v(B) = -\langle \phi_v(r) | V_c(\vec{r})\ \Delta_H | \phi_v(r) \rangle = -\langle \phi_v(r) | \delta\gamma V_c(\vec{r}) | \phi_v(r) \rangle, \tag{7}$$

where $Z$ represents the atomic charge number, $\gamma = 1 + \delta\gamma$ are the binding-energy ratio and relative binding-energy ratio $\delta\gamma$, respectively, and $B$ indicates the bulk atoms. $\varepsilon_0$ is the dielectric constant of the vacuum.

The energy-level shifts in an external field obtained from formulas 5, 6 and 7 are:

$$\begin{cases} \delta\gamma = -1(Antibonding),\ \Delta E_v(x) = \Delta E_v(0) = 0 & \text{(isolated atom,neutral atom)} \\ \delta\gamma = 0(Bonding), \Delta E_v(x) = \Delta E_v(B) = -(-\sum\limits_{l,\vec{R}_l \neq 0} \langle \phi_v(r) | \frac{1}{4\pi\varepsilon_0} \frac{Z'e^2}{\left|\vec{r} - \vec{R}_l\right|} | \phi_v(r) \rangle) > 0 & \text{(bulk atoms,core level loses electrons)} \\ \delta\gamma > 0 \begin{cases} \delta\gamma > 1(Bonding), \Delta E_v(x) = -(-\langle \phi_v(r) | \sum\limits_{l,\vec{R}_l \neq 0} \frac{(1+\delta\gamma)}{4\pi\varepsilon_0} \frac{Z'e^2}{\left|\vec{r} - \vec{R}_j\right|} | \phi_v(r) \rangle) > 0 & \text{(potential well becomes deeper)} \\ 1 > \delta\gamma > 0(Bonding),\ \Delta E_v(x) = -(-\langle \phi_v(r) | \sum\limits_{l,\vec{R}_l \neq 0} \frac{(1+\delta\gamma)}{4\pi\varepsilon_0} \frac{Z'e^2}{\left|\vec{r} - \vec{R}_j\right|} | \phi_v(r) \rangle) > 0 & \text{(core level loses electrons)} \end{cases} \\ \delta\gamma < 0 \begin{cases} -1 < \delta\gamma < 0(Nonbonding),\ \Delta E_v(x) = -(-\sum\limits_{l,\vec{R}_l \neq 0} \langle \phi_v(r) | \frac{(1+\delta\gamma)}{4\pi\varepsilon_0} \frac{Z'e^2}{\left|\vec{r} - \vec{R}_j\right|} | \phi_v(r) \rangle) > 0 & \text{(core level gets electrons)} \\ \delta\gamma < -1(Antibonding), \Delta E_v(x) = -(-\sum\limits_{l,\vec{R}_l \neq 0} \langle \phi_v(r) | \frac{(1+\delta\gamma)}{4\pi\varepsilon_0} \frac{Z'e^2}{\left|\vec{r} - \vec{R}_j\right|} | \phi_v(r) \rangle < 0) & \text{( potential well becomes shallower or barrier potential)} \end{cases} \end{cases} \tag{8}$$

Formula 8 was obtained by calculating the external field-induced binding-energy (BE)

ratio, $\gamma$, using the known reference values of $\Delta E_v'(x) = E_v(x) - E_v(B)$, $\Delta E_v(B) = E_v(B) - E_v(0)$, and $\Delta E_v(x) = E_v(x) - E_v(0)$ derived from X-ray photoelectron spectroscopy (XPS) analysis[28]. Then, we obtain

$$\gamma = \frac{E_v(x) - E_v(0)}{E_v(B) - E_v(0)} \approx \frac{Z_x}{Z_b}\frac{d_b}{d_x} = \left(\frac{Z_b - \sigma'}{Z_b}\right)\left(\frac{d_x}{d_b}\right)^{-1} = \left(\frac{d_x}{d_b}\right)^{-m} = \frac{E_x}{E_b}, \tag{9}$$

$\sigma'$ is the relative charge shielding factor. $m$ is an indicator for the bond nature of a specific material. The $m$ is related to the charge shielding factor $\sigma'$,

$m = 1 - \dfrac{In \frac{Z_b - \sigma'}{Z_b}}{In\left(\frac{d_x}{d_b}\right)}$. The binding energy shifts in the BBC model is derived from the extended model of Bond–Order–Length–Strength (BOLS) theory. The difference is that we use the central force field method and considering the effect of electronic shielding, the DFT calculation data is basically consistent with the XPS experimental data [29].

The Formula 10 was obtained by considering the effect of the charge transfer of the bond-charge model:

$$\begin{aligned} E[\rho_0+\delta\rho] &= E^0[\rho_0] + E^1[\rho_0+\delta\rho] + E^2\left[\rho_0,(\delta\rho)^2\right] \\ &\approx \left(\sum_i \langle\phi_i| -\frac{1}{2}\nabla^2 + V_{eff}(\vec{r})|\phi_i\rangle - \frac{1}{2}\iint \frac{\rho_0(\vec{r})\rho_0(\vec{r}')}{|\vec{r}-\vec{r}'|}\mathrm{d}\vec{r}\mathrm{d}\vec{r}' - \int V^{XC}[\rho_0]\rho_0(\vec{r})\mathrm{d}\vec{r} + E^{XC}[\rho_0]\right) + \sum_i f_i\varepsilon_i + \frac{1}{2}\iint \frac{1}{|\vec{r}-\vec{r}'|}\delta\rho(\vec{r})\delta\rho(\vec{r}')\mathrm{d}\vec{r}\mathrm{d}\vec{r}' \end{aligned} \tag{10}$$

where $E^0[\rho_0]$ is termed the "repulsive energy" which determines the dispersion of the energy band. $V^{XC}$ and $E^{XC}$ are the potential and exchange correlation energies, respectively, and are typically fitted from the DFT calculations. In the second term, $E^1[\rho_0+\delta\rho]$, $\varepsilon_i$ indicates the $i$th electronic level and $f_i$ is the corresponding electronic occupation number. In the second term, $E^2\left[\rho_0,(\delta\rho)^2\right]$, the $\rho_0(\vec{r})$ is the initial density function, and $\delta\rho(\vec{r})$ is the deformation density function. Thereafter, formula 9 was obtained:

$$\Delta V_{bc}(\vec{r}-\vec{r}') = \frac{1}{8\pi\varepsilon_0}\int d^3r\int d^3r' \frac{\delta\rho(\vec{r})\delta\rho(\vec{r}')}{|\vec{r}-\vec{r}'|},$$

(11)

where $\Delta V_{bc}(\vec{r}_{ij})$ is the deformation charge of the bond energy. Additionally, the deformation density $\delta\rho$ satisfies the following relationship (formula 12):

$$(\delta\rho_{Hole\text{-}electron} \le \delta\rho_{Antibonding\text{-}electron} < \delta\rho_{No\ charge\ tranfer} = 0 < \delta\rho_{Nonbonding\text{-}electron} \le \delta\rho_{Bonding\text{-}electron})$$

(12)

For bonding (formula 13):

$$\delta\rho_{Hole\text{-}electron}(\vec{r})\,\delta\rho_{Bonding\text{-}electron}(\vec{r}') < 0(Strong\ Bonding).$$

(13)

For nonbonding or weak bonding (formula 14):

$$\begin{cases} \delta\rho_{Hole\text{-}electron}(\vec{r})\,\delta\rho_{Nonbonding\text{-}electron}(\vec{r}') < 0(Nonbonding\ or\ Weak\ Bonding) \\ \delta\rho_{Antibonding\text{-}electron}(\vec{r})\,\delta\rho_{Bonding\text{-}electron}(\vec{r}') < 0(Nonbonding\ or\ Weak\ Bonding) \\ \delta\rho_{Antibonding\text{-}electron}(\vec{r})\,\delta\rho_{Nonbonding\text{-}electron}(\vec{r}') < 0(Nonbonding) \end{cases}$$

(14)

For antibonding (formula 15):

$$\begin{cases} \delta\rho_{Nonbonding\text{-}electron}\left(\vec{r}\right)\delta\rho_{Bonding\text{-}electron}\left(\vec{r}'\right) > 0(Antibonding) \\ \delta\rho_{Hole\text{-}electron}\left(\vec{r}\right)\delta\rho_{Antibonding\text{-}electron}\left(\vec{r}'\right) > 0(Antibonding) \\ \delta\rho_{Hole\text{-}electron}\left(\vec{r}\right)\delta\rho_{Hole\text{-}electron}\left(\vec{r}'\right) > 0(Antibonding) \\ \delta\rho_{Antibonding\text{-}electron}\left(\vec{r}\right)\delta\rho_{Antibonding\text{-}electron}\left(\vec{r}'\right) > 0(Antibonding) \\ \delta\rho_{Nonbonding\text{-}electron}\left(\vec{r}\right)\delta\rho_{Nonbonding\text{-}electron}\left(\vec{r}'\right) > 0(Antibonding) \\ \delta\rho_{Bonding\text{-}electron}\left(\vec{r}\right)\delta\rho_{Bonding\text{-}electron}\left(\vec{r}'\right) > 0(Antibonding) \end{cases}$$

(15)

It is noteworthy that the formation of the chemical bonds is related to the fluctuations in electron density. For the total deformation charge of the bond energy, we can write it as:

$$\begin{aligned} \Delta\overline{V}_{bc}(\vec{r}-\vec{r}') &= \frac{1}{N}\left(\sum_i V_{att}(\vec{r}-\vec{r}') + \sum_j V_{rep}(\vec{r}-\vec{r}')\right) \\ &= \frac{1}{N}\left(\Delta\hat{V}_{Bonding}(\vec{r}-\vec{r}') + \Delta\hat{V}_{Nonbonding}(\vec{r}-\vec{r}') + \Delta\hat{V}_{Antibonding}(\vec{r}-\vec{r}')\right) \quad (N = i + j) \end{aligned}$$

(16)

$V_{rep}(\vec{r}-\vec{r}')$ represents the bond energy of repulsion and $V_{att}(\vec{r}-\vec{r}')$ represents the bond energy of attraction.

Consider the influence of external field on deformation charge of the bond energy, we introduce the scale of bond energy:

$$\Delta\overline{V}_{bci}(\vec{r}-\vec{r}') = \Delta\overline{V}_{bc0}(\vec{r}-\vec{r}')\mathrm{e}^{-\lambda r_s}$$

(17)

$\lambda$ is the external field shielding parameter of bond energy deformation charge.

From the Hubbard model, we get

$$\hat{V}_{ee} = \frac{1}{2}\int d^3r \int d^3r' \alpha_{\zeta}{}^{+}(\vec{r})\alpha_{\zeta}(\vec{r})V_{ee}(\vec{r}-\vec{r}')\alpha_{\zeta'}{}^{+}(\vec{r}')\alpha_{\zeta'}(\vec{r}')$$
$$= \frac{1}{2|\vec{r}-\vec{r}'|}\int d^3r \int d^3r' \rho(\vec{r})\rho(\vec{r}')$$

(18)

where $\alpha_{\zeta}(\vec{r})$ and $\alpha_{\zeta'}{}^{+}(\vec{r})$ are annihilation and creation operators, respectively. For the sake of completeness, we have endowed the electrons with a spin index, $\zeta = \uparrow / \downarrow$. The charge density is $\rho(\vec{r}) = \alpha_{\zeta}^{+}(\vec{r})\alpha_{\zeta}^{-}(\vec{r})$. The quantity $V_{ee}(\vec{r}-\vec{r}') = \frac{1}{2|\vec{r}_i - \vec{r}_j'|}$ is the potential of the electron, which induces a transformation

$$\alpha_{\zeta}^{+}(\vec{r}) = \sum_{\vec{R}} \psi_{\vec{R}}^{*}(\vec{r}) a_{\vec{R}\zeta}^{+} \equiv \sum_{i} \psi_{\vec{R}i}^{*}(\vec{r}) a_{i\zeta}^{+} \quad .$$

(19)

Inserting (19) into (18) leads to the expansion $\hat{V}_{\text{ee}} = \sum_{ii'jj'} U_{ii'jj'} a_{i\zeta}^{+} a_{i'\zeta'}^{+} a_{j\zeta}^{-} a_{j'\zeta'}^{-}$, where

$$U_{ii'jj'} = \frac{1}{2}\int d^3r \int d^3r' \psi_{\vec{R}i}^{*}(\vec{r})\psi_{\vec{R}j}(\vec{r})V_{ee}(\vec{r}-\vec{r}')\psi_{\vec{R}i'}^{*}(\vec{r}')\psi_{\vec{R}j'}(\vec{r}')$$

(20)

is the Coulomb interaction.

Therefore, the electron density fluctuations can also be represented by the Hubbard model.

$$\Delta V_{bc}(\vec{r}-\vec{r}') = \pm\frac{1}{2}\int d^3r \int d^3r' \alpha_{\zeta}{}^{+}(\vec{r})\alpha_{\zeta}(\vec{r})V_{ee}(\vec{r}-\vec{r}')\alpha_{\zeta'}{}^{+}(\vec{r}')\alpha_{\zeta'}(\vec{r}')$$
$$= \frac{1}{2|\vec{r}-\vec{r}'|}\int d^3r \int d^3r' \delta\rho(\vec{r})\delta\rho(\vec{r}')$$

(21)

The deformation charge-bond energy $\Delta V_{bc}(\vec{r}-\vec{r}')$ is different from the Coulomb repulsion energy $\hat{V}_{ee}$. The deformation charge-bond energy considers the interaction between electrons and holes, and there are three cases of bonding, antibonding and nonbonding. For the bond-charge model of BBC model, its source is the second order term of energy expansion. The formula is linked to the BBC model, which can be found in **supporting materials**.

### 2.3 Time dependent first-order response wave function

For systems of linear differential equations with constant coefficients：

$$x'(t) = Ax(t) \tag{22}$$

*A* is a matrix.Column vector *x* is the physical quantity of first-order time response that can be derived.For example, wave function $\varphi$ , electric field *E*, magnetic field *B* and density $\rho$ , etc.

Consider homogeneous linear differential equations with constant coefficients of two variables

$$\begin{cases} \dfrac{d}{dt}\begin{pmatrix} x_1(t) \\ x_2(t) \end{pmatrix} = A\begin{pmatrix} x_1(t) \\ x_2(t) \end{pmatrix} \\ A = \begin{pmatrix} c_{11} & c_{12} \\ c_{21} & c_{22} \end{pmatrix} \end{cases} \tag{23}$$

Let $\lambda_1$ and $\lambda_2$ be the characteristic roots of coefficient matrix A of the equation (23), then

（i）When $c_{12}=0,\ c_{11}=c_{22}$, the basic solution matrix of equation (23) is

$$\Phi(t) = \begin{bmatrix} e^{c_{11}t} & 0 \\ a_{21}te^{c_{11}t} & e^{c_{11}t} \end{bmatrix}$$

(24)

（ii） When $c_{12}=0,\ c_{11}\neq c_{22}$, the basic solution matrix of equation (23) is

$$\Phi(t)=\begin{bmatrix} e^{c_{11}t} & 0 \\ \dfrac{c_{21}}{c_{11}-c_{22}}\cdot e^{c_{11}t} & e^{c_{22}t} \end{bmatrix}$$

(25)

（iii）When $c_{12}\neq 0$, and $\lambda_1,\ \lambda_2$ are unequal real roots, the basic solution matrix of equation (23) is

$$\Phi(t)=\begin{bmatrix} e^{\lambda_1 t} & e^{\lambda_2 t} \\ \dfrac{\lambda_1-c_{11}}{c_{12}}e^{\lambda_1 t} & \dfrac{\lambda_2-c_{11}}{c_{12}}e^{\lambda_2 t} \end{bmatrix}$$

(26)

(iv) When $c_{12}\neq 0$, and $\lambda_1,\ \lambda_2$ are equal real roots, the basic solution matrix of equation (23) is

$$\Phi(t)=\begin{bmatrix} e^{\lambda_1 t} & te^{\lambda_1 t} \\ \dfrac{\lambda_1-c_{11}}{c_{12}}e^{\lambda_1 t} & \dfrac{1+(\lambda_1-c_{11})t}{c_{12}}e^{\lambda_1 t} \end{bmatrix}$$

(27)

(v) When $c_{12}\neq 0$, and $\lambda_1,\ \lambda_2$ are conjugate complex roots, the basic solution matrix of equation (23) is

$$\Phi(t)=\begin{bmatrix} e^{\alpha t}\cos\beta t & e^{\alpha t}\sin\beta t \\ e^{at}\cdot\dfrac{(\alpha-c_{11})\cos\beta t-\beta\sin\beta t}{c_{12}} & e^{at}\cdot\dfrac{(\alpha-c_{11})\sin\beta t-\beta\cos\beta t}{c_{12}} \end{bmatrix}$$

(28)

Consider wave function of *sp* equivalent hybridization:

$$\begin{cases} \varphi_1 = \sqrt{\frac{1}{2}}\varphi_s + \sqrt{\frac{1}{2}}\varphi_p \\ \varphi_2 = \sqrt{\frac{1}{2}}\varphi_s - \sqrt{\frac{1}{2}}\varphi_p \end{cases}$$

Then, matrix formation:

(29)

$$\begin{pmatrix} \varphi_1 \\ \varphi_2 \end{pmatrix} = \begin{pmatrix} \sqrt{\frac{1}{2}} & \sqrt{\frac{1}{2}} \\ \sqrt{\frac{1}{2}} & -\sqrt{\frac{1}{2}} \end{pmatrix} \begin{pmatrix} \varphi_s \\ \varphi_p \end{pmatrix}$$

(30)

If the wave function obtained is a first-order response function with time,we have:

$$\frac{d}{dt}\begin{pmatrix} \varphi_1(t) \\ \varphi_2(t) \end{pmatrix} = \begin{pmatrix} \sqrt{\frac{1}{2}} & \sqrt{\frac{1}{2}} \\ \sqrt{\frac{1}{2}} & -\sqrt{\frac{1}{2}} \end{pmatrix} \begin{pmatrix} \varphi_s(t) \\ \varphi_p(t) \end{pmatrix}$$

(31)

The eigenvalue of the equation are $\lambda_1 = 1$ and $\lambda_2 = -1$.

$$\Phi(t) = \begin{bmatrix} e^t & e^{-t} \\ \frac{1-\sqrt{\frac{1}{2}}}{\sqrt{\frac{1}{2}}} e^t & \frac{-1-\sqrt{\frac{1}{2}}}{\sqrt{\frac{1}{2}}} e^{-t} \end{bmatrix}$$

(32)

According to Hamilton cayley theorem, $p(\lambda)$ is the characteristic polynomial of the matrix, then

$$p(A)=A^n+a_1A^{n-1}+\cdots+a_{n-1}A+a_nE=0,$$

Or

$$p(A)=(A-\lambda_nE)(A-\lambda_{n-1}E)\cdots(A-\lambda_1E)=0.$$

(33)

$\lambda_1,\lambda_2,\cdots,\lambda_n$ are the *n* eigenvalues of matrix A (they are not necessarily equal), then

$$\exp At=\sum_{i=0}^{n} r_{i+1}(t)P_i$$

and

$P_0=E, P_i=(A-\lambda_iE)(A-\lambda_{i-1}E)\cdots(A-\lambda_1E)\quad (i=1,2,\cdots,n)$, $r_i(t)(i=1,2,\cdots,n)$ is a system of equations

$$\begin{cases} r_1'(t)=\lambda_1 r_1(t), \\ r_{i+1}'(t)=r_i(t)+\lambda_{i+1}r_{i+1}(t) \quad (i=1,2,\cdots,n-2) \\ r_n'(t)=r_{n-1}(t)+\lambda_n r_n(t) \end{cases}$$

(34)

Solution satisfying initial condition

$$r_1(0)=1, r_2(0)=0,\cdots,r_n(0)=0$$

(35)

Consider wave function of $sp^2$ equivalent hybridization:

$$\begin{cases} \varphi_1=\sqrt{\frac{1}{2}}\varphi_s+\sqrt{\frac{2}{3}}\varphi_{px} \\ \varphi_2=\sqrt{\frac{1}{3}}\varphi_s-\sqrt{\frac{1}{6}}\varphi_{px}+\sqrt{\frac{1}{2}}\varphi_{py} \\ \varphi_3=\sqrt{\frac{1}{3}}\varphi_s-\sqrt{\frac{1}{6}}\varphi_{px}-\sqrt{\frac{1}{2}}\varphi_{py} \end{cases}$$

(36)

Then, matrix formation:

$$\begin{pmatrix} \varphi_1 \\ \varphi_2 \\ \varphi_3 \end{pmatrix} = \begin{pmatrix} \sqrt{\frac{1}{2}} & \sqrt{\frac{2}{3}} & 0 \\ \sqrt{\frac{1}{3}} & -\sqrt{\frac{1}{6}} & \sqrt{\frac{1}{2}} \\ \sqrt{\frac{1}{3}} & -\sqrt{\frac{1}{6}} & -\sqrt{\frac{1}{2}} \end{pmatrix} \begin{pmatrix} \varphi_s \\ \varphi_{px} \\ \varphi_{py} \end{pmatrix}$$

(37)

If the wave function obtained is a first-order response function with time,we have:

$$\frac{d}{dt}\begin{pmatrix} \varphi_1(t) \\ \varphi_2(t) \\ \varphi_3(t) \end{pmatrix} = \begin{pmatrix} \sqrt{\frac{1}{2}} & \sqrt{\frac{2}{3}} & 0 \\ \sqrt{\frac{1}{3}} & -\sqrt{\frac{1}{6}} & \sqrt{\frac{1}{2}} \\ \sqrt{\frac{1}{3}} & -\sqrt{\frac{1}{6}} & -\sqrt{\frac{1}{2}} \end{pmatrix} \begin{pmatrix} \varphi_s(t) \\ \varphi_{px}(t) \\ \varphi_{py}(t) \end{pmatrix}$$

(38)

The eigenvalue of the equation are $\lambda_1 = 1$ , $\lambda_2 \approx -0.769 - 0.639\,i$ and $\lambda_3 \approx -0.769 + 0.639\,i$ .

Solve the initial value:

$$\begin{cases} r_1'(t) = r_1(t) \\ r_2'(t) = r_1(t) - (0.769 + 0.639\,i)r_2(t) \\ r_3'(t) = r_2(t) - (0.769 - 0.639\,i)r_3(t) \\ r_1(0) = 1,\ r_2(0) = r_3(0) = 0 \end{cases}$$

(39)

First solve the initial value:

$$\begin{cases} r_1'(t) = e^t \\ r_1(0) = 1 \end{cases}$$

(40)

The solution is as follows:

$$r_1(t) = e^t \tag{41}$$

Solve the initial value again:

$$\begin{cases} r_2{}'(t) \approx e^t - (0.769 + 0.639\, i) r_2(t) \\ r_2(0) = 0 \end{cases} \tag{42}$$

The solution is as follows：

$$r_2(t) \approx (0.500 - 0.181i)e^t - (0.500 - 0.181i)e^{(-0.769 - 0.639\, i)\, t} \tag{43}$$

After that, solve the initial value:

$$\begin{cases} r_3{}'(t) \approx (0.500 - 0.181i)e^t - (0.500 - 0.181i)e^{(-0.769 - 0.639\, i)\, t} - (0.769 - 0.639\, i) r_3(t) \\ r_3(0) = 0 \end{cases} \tag{44}$$

The solution is as follows：

$$r_3(t) \approx e^{(-0.769 - 1.278i)\, t}(-(0.142 + 0.391i)e^t + 0.283 e^{(1.769 - 0.639i)\, t}) + (0.391i - 0.141)\, e^{(-0.769 + 0.639\, i)\, t} \tag{45}$$

The basic solution matrix of the equation system is:

$$\begin{aligned} \Phi(t) &= r_1(t)\, P + r_2(t)\, P_1 + r_3(t)\, P_2 \\ &= r_1(t)E + r_2(t)\left(\mathrm{A} - E\right) + r_3(t)\left(\mathrm{A} - E\right)\left(\mathrm{A} - (-0.769 - 0.639\, i)E\right) \end{aligned} \tag{46}$$

## 3. Results and discussion

### 3.1 Geometric structure of $Cd_{43}Te_{28}$ microporous material

We established the geometric structure of the $Cd_nTe_m$ nanocluster for the initial structure. Thereafter, the nanocluster was optimized by adjusting the supercell size to obtain the **$Cd_{43}Te_{28}$** micropore structure. The structures of $Cd_{43}Te_{28}$ is illustrated in **Fig. 2**. The optimized results of the lattice parameters and atomic positions of microporous $Cd_{43}Te_{28}$ is listed in **Table 1**. The lattice parameters share the same a, b, and c (16.118 Å) values for $Cd_{43}Te_{28}$, and the atomic positions are all 90.00° in *α*, *β*, and *γ*. Molecular dynamics simulation shows that $Cd_{43}Te_{28}$ is stable. The results are shown in the supplemental material of **Table S1** and **Figs. S1-S2**. In addition, the optimized atomic coordinates are listed in **Table 3**, which lists the physical and chemical properties of microporous $Cd_{43}Te_{28}$ for subsequent research. The average diameter for the micropores of $Cd_{43}Te_{28}$ is 8.289 Å (~0.83 nm), respectively. A microporous material is a material containing pores with diameters less than 2 nm. For example, ETS-10 is a synthetic titanosilicate with a three-dimensional 12-ring channel system with 0.49×0.76 nm micropores. [30]

### 3.2 Band structure of $Cd_{43}Te_{28}$ microporous material

The energy band values of $Cd_{43}Te_{28}$ microporous materials calculated by PBE and LDA potentials are 0.835eV and 0.799 eV, respectively. The calculated energy band value may be between 0.799 ~ 0.835 eV. The delocalization is weak when the energy distribution is smooth, and the density is high [31]. The band gap width is narrow in these structures, and there is no intersection point between the valence and conduction bands. The 5*p* orbital of Te has four electrons, and the *p* energy level can accommodate six electrons, in theory. The 5*s* orbital of the Cd has two electrons, and the *s* orbital is filled. Overall, the $Cd_nTe_m$ microporous structure semiconductor has empty orbitals and is mainly hole conductive. This phenomenon occurs because the Fermi level is closer to the highest point of the valence band. Hence, $Cd_nTe_m$ can be considered a *p*-type semiconductor [32]. Semiconductors with this type of microporous structure may exhibit an excellent catalytic performance. For example: Li et al. found that Microporous organic polymers (MOPs) as heterogeneous photocatalysts for visible light promotion [33].

### 3.3 Density of states of $Cd_{43}Te_{28}$ microporous material

In **Fig. 3**, the DOS(density of states) diagrams of $Cd_{43}Te_{28}$ was given. The main peaks of the electron density distribution occur at -1.52 eV for $Cd_{43}Te_{28}$. The structure is relatively stable because of the main peaks is located in the direction of negative energy [34]. Moreover, the zero field can be observed near the Fermi level, and the $Cd_{43}Te_{28}$ status as a semiconductor has been confirmed from the energy band diagram. The energy region of the three structures fluctuates significantly and is typical of a *p*-type band structure. The *p* electrons are relatively localized, which corresponds with the narrow band gap [35]. The partial density of states (PDOS) was analyzed for $Cd_{43}Te_{28}$. **Fig. 3** illustrates that the change in the $Cd_{43}Te_{28}$ structure DOS curve is the largest, and Cd contributes substantially to the $Cd_{43}Te_{28}$ structure. The contribution of Cd was confirmed by the presence of obvious peaks throughout the energy range.

### 3.4 Deformation charge density and bonding of $Cd_{43}Te_{28}$ microporous material

The deformation charge density is calculated by DFT and can be obtained via contribution from four bonding and electronic features. The deformation charge densities of $Cd_{43}Te_{28}$ are shown in **Fig. 4.** The deformation charge density scale indicates the charge value. The red and blue are as represent the increase and decrease of the electron distribution respectively. The electrons in the positive region are convergent, while they are divergent in the negative region [36]. The red and white regions represent numerous electrons gathering, and the deformation charge density is positive [37] .The established BBC model was used to convert the value of Hamiltonian into the value of bonding (**Table 4**). Using Formula **11**, the deformation bond energies is -0.0797 eV for $Cd_{43}Te_{28}$.We have developed a method to accurately analyze the chemical bonding and electronic properties of microporous semiconductor materials. In conclusion, our calculations provide a theoretical reference for the bonding properties of $Cd_nTe_m$ microporous semiconductor materials.

### 3.5 Geometric topological bonding and wave function

By introducing the phase into the chemical bond, the topological bonding information is obtained:

（1）Take *z* as the vertical direction:

$$\begin{cases} f(x_i) = x_i + \left(r_i e^{-\chi r_i} + \mathrm{ASin}[v]\right) * \mathrm{Cos}[u] \\ f(y_i) = y_i + \left(r_i e^{-\chi r_i} + \mathrm{ASin}[v]\right) * \mathrm{Sin}[u] \\ f(z_i) = z_i + \mathrm{BCos}[v] \end{cases} \tag{47}$$

（2）Take *y* as the vertical direction:

$$\begin{cases} f(x_i) = x_i + \left(r_i e^{-\chi r_i} + \mathrm{ASin}[v]\right) * \mathrm{Cos}[u] \\ f(y_i) = y_i + \mathrm{BCos}[v] \\ f(z_i) = z_i + \left(r_i e^{-\chi r_i} + \mathrm{ASin}[v]\right) * \mathrm{Sin}[u] \end{cases} \tag{48}$$

（3）Take $x$ as the vertical direction:

$$\begin{cases} f(x_i) = x_i + \text{BCos}[v] \\ f(y_i) = y_i + \left(r_i e^{-\chi r_i} + \text{ASin}[v]\right) * \text{Sin}[u] \\ f(z_i) = z_i + \left(r_i e^{-\chi r_i} + \text{ASin}[v]\right) * \text{Cos}[u] \end{cases}$$

(49)

The $x_i$, $y_i$ and $z_i$ represent the position coordinates of the atom, $r_i$ represent the radius of the atom, $\chi$ is electron shielding parameter, A and B represent the amplitude of the wave, and $u$ and $v$ represent the phase of the wave.We use Formula **47-49** to calculate the geometric topological bonding of $Cd_{43}Te_{28}$ microscopic semiconductor materials, as is shown in **Fig. 5 .** The calculated parameters are shown in **Table 5**. The Formula 47-49 can obtain the topological form of atoms, which is related to the experimental chemical reaction dynamic [38].

In section 2.3, we discuss the time dependent first-order response wave function. For topological chemical bonds, we need to use topological wave function. We call magic cube(Rubik's Cube) [39]wave function, as is shown in the **Fig. 6**. **Fig. 7** shows the operation of the magic cube matrix wave function. $F_{m(n)}$, $B_{m(n)}$, $L_{m(n)}$, $R_{m(n)}$, $U_{m(n)}$ and $D_{m(n)}$ are rotation operators of magic cube matrix. $F'_{m(n)}$, $B'_{m(n)}$, $L'_{m(n)}$, $R'_{m(n)}$, $U'_{m(n)}$ and $D'_{m(n)}$ are the rotation operators of the inverse of the magic cube matrix. The subscript m represents the number of revolutions. A letter by itself (e.g. F) means turn that face 90 degrees clockwise with respect to the center of the cube. A letter with an apostrophe (F') denotes a 90 degree counter-clockwise turn. A letter followed by the number 2 ($F_2$) denotes 2 turns, i.e. a 180 degree turn. The subscript n indicates the number of operations. **In Fig. 7**, we use the rotation operators of the magic cube matrix wave function to write:

$$\begin{cases} \left\langle R_{1(1)} U_{1(2)} R'_{1(3)} \left\| \begin{matrix} \psi_{11} & \psi_{12} & \psi_{13} \\ \psi_{21} & \psi_{22} & \psi_{23} \\ \psi_{31} & \psi_{32} & \psi_{33} \end{matrix} \right| \right. = \left\langle R_{1(1)} U_{1(2)} \left\| \begin{matrix} \psi_{11} & \psi_{12} & \psi_{13} \\ \psi_{21} & \psi_{22} & \psi_{23} \\ \psi_{31} & \psi_{32} & \psi_{33} \end{matrix} \right\| R_{1(3)} \right\rangle \\ \left\langle R'_{1(3)} \right| = \left| R_{1(3)} \right\rangle \end{cases}$$

(50)

Formula 50 gives the matrix transformation process of the magic cube wave function operated by the rotation operator.

## 4. Conclusions

In this study, the band structure, DOS, and deformation charge density of $Cd_{43}Te_{28}$ microporous materials was determined using DFT calculations. The band gap values of $Cd_{43}Te_{28}$ structures is 0.835 eV. For semiconductor microporous materials, it is necessary to regulate the physical and chemical properties of a novel structure from the perspective of semiconductor application. In addition, the established BBC model was used to convert the value of Hamiltonian into the bonding value. We provide a method for describing topological chemical bonds by atomic coordinates and wave phases. We also discuss the dynamic process of the wave function with time and the magic cube matrix. This work provides an innovative method and technology for the accurate analysis of the bond and electronic properties for microporous semiconductor materials.

**Acknowledgements**

The Chongqing Natural Science Foundation project (cstc2019jcyj-msxmX0674)

## References

[1] J. Newsam, M. M. Treacy, W. Koetsier, C. d. Gruyter *Proceedings of the royal society of London. A. mathematical and physical sciences*. **1988**, *420*, 375-405.
[2] R. J. Francis, D. O'Hare *Journal of the Chemical Society, Dalton Transactions*. **1998**, 3133-3148.
[3] J. Li, A. Corma, J. Yu *Chemical Society Reviews*. **2015**, *44*, 7112-7127.
[4] S. M. Csicsery *Zeolites*. **1984**, *4*, 202-213.
[5] W. W. Yu, L. Qu, W. Guo, X. Peng *Chemistry of materials*. **2003**, *15*, 2854-2860.
[6] S. Kim, B. Fisher, H.-J. Eisler, M. Bawendi *Journal of the American Chemical Society*. **2003**, *125*, 11466-11467.
[7] J. H. Bang, P. V. Kamat *ACS nano*. **2009**, *3*, 1467-1476.
[8] I. Dharmadasa, A. Ojo *Journal of Materials Science: Materials in Electronics*. **2017**, *28*, 16598-16617.
[9] A. Ukarande, S. Chaure, N. B. Chaure *Materials Today: Proceedings*. **2022**, *68*, 2683-2686.
[10] A. Bosio, N. Romeo, S. Mazzamuto, V. Canevari *Progress in Crystal Growth and Characterization of Materials*. **2006**, *52*, 247-279.
[11] X. Wu *Solar energy*. **2004**, *77*, 803-814.
[12] M. Gao, S. Kirstein, H. Möhwald, A. L. Rogach, A. Kornowski, A. Eychmüller, H. Weller *The Journal of Physical Chemistry B*. **1998**, *102*, 8360-8363.
[13] S. J. Cho, D. Maysinger, M. Jain, B. Röder, S. Hackbarth, F. M. Winnik *Langmuir*. **2007**, *23*, 1974-1980.

[14] Z. Tang, N. A. Kotov, M. Giersig *Science*. **2002**, *297*, 237-240.
[15] H. Qiu, H. Li, J. Wang, Y. Zhu, M. Bo *physica status solidi (RRL)–Rapid Research Letters*. **2022**, *16*, 2100444.
[16] M. Bo, H. Li, Z. Huang, L. Li, C. Yao *AIP Advances*. **2020**, *10*, 015321.
[17] M. Bo, H. Li, A. Deng, L. Li, C. Yao, Z. Huang, C. Peng *Materials Advances*. **2020**, *1*, 1186-1192.
[18] K. Funabashi, J. L. Magee *The Journal of Chemical Physics*. **1957**, *26*, 407-411.
[19] W. M. C. Foulkes, R. Haydock *Physical review B*. **1989**, *39*, 12520.
[20] R. Y. Chiao, Y.-S. Wu *Physical review letters*. **1986**, *57*, 933.
[21] M. Segall, P. J. Lindan, M. a. Probert, C. J. Pickard, P. J. Hasnip, S. Clark, M. Payne *Journal of physics: condensed matter*. **2002**, *14*, 2717.
[22] B. Hammer, L. B. Hansen, J. K. Nørskov *Physical review B*. **1999**, *59*, 7413.
[23] X. Huang, B. Li, C. Peng, G. Song, Y. Peng, Z. Xiao, X. Liu, J. Yang, L. Yu, J. Hu *Nanoscale*. **2016**, *8*, 1040-1048.
[24] D. C. Langreth, M. Mehl *Physical Review B*. **1983**, *28*, 1809.
[25] X. Liu, X. Zhang, M. Bo, L. Li, H. Tian, Y. Nie, Y. Sun, S. Xu, Y. Wang, W. Zheng *Chemical reviews*. **2015**, *115*, 6746-6810.
[26] C. Q. Sun *Physical review B*. **2004**, *69*, 045105.
[27] M. A. Omar, Elementary solid state physics: principles and applications, Pearson Education India, **1975**.
[28] C. Q. Sun, Relaxation of the chemical bond: skin chemisorption size matter ZTP mechanics H2O myths, Springer, **2014**.
[29] W. Zhu, Z. Huang, M. Bo, J. Huang, C. Peng, H. Liu *Chinese Journal of Chemical Physics*. **2021**, *34*, 628.
[30] Z. Lin, J. Rocha in *Crystallization of microporous titanosilicate membranes from clear solutions, Vol. 170*, Elsevier, **2007**, pp.493-498.
[31] Q. Deng, F. Zhou, M. Bo, Y. Feng, Y. Huang, C. Peng *Applied Surface Science*. **2021**, *545*, 149008.
[32] S. Levytskyi, T. Zhao, Z. Cao, A. Stronski *Physics and Chemistry of Solid State*. **2021**, *22*, 301-306.
[33] R. Li, Z. J. Wang, L. Wang, B. C. Ma, S. Ghasimi, H. Lu, K. Landfester, K. A. Zhang *Acs Catalysis*. **2016**, *6*, 1113-1121.
[34] M. Irfan, J. Iqbal, S. Sadaf, B. Eliasson, U. A. Rana, S. Ud-din Khan, K. Ayub *International Journal of Quantum Chemistry*. **2017**, *117*, e25363.
[35] I. Pollini *Physical Review B*. **1996**, *53*, 12769.
[36] Y. Feng, F. Zhou, M. Bo, Y. Huang, Q. Deng, C. Peng *The Journal of Physical Chemistry C*. **2020**, *124*, 27780-27789.
[37] R. Meng, Q. Deng, C. Peng, B. Chen, K. Liao, L. Li, Z. Yang, D. Yang, L. Zheng, C. Zhang *Nano Today*. **2020**, *35*, 100991.
[38] Y. Xie, H. Zhao, Y. Wang, Y. Huang, T. Wang, X. Xu, C. Xiao, Z. Sun, D. H. Zhang, X. Yang *Science*. **2020**, *368*, 767-771.
[39] F. Agostinelli, S. McAleer, A. Shmakov, P. Baldi *Nature Machine Intelligence*. **2019**, *1*, 356-363.

**Figures and Tables**

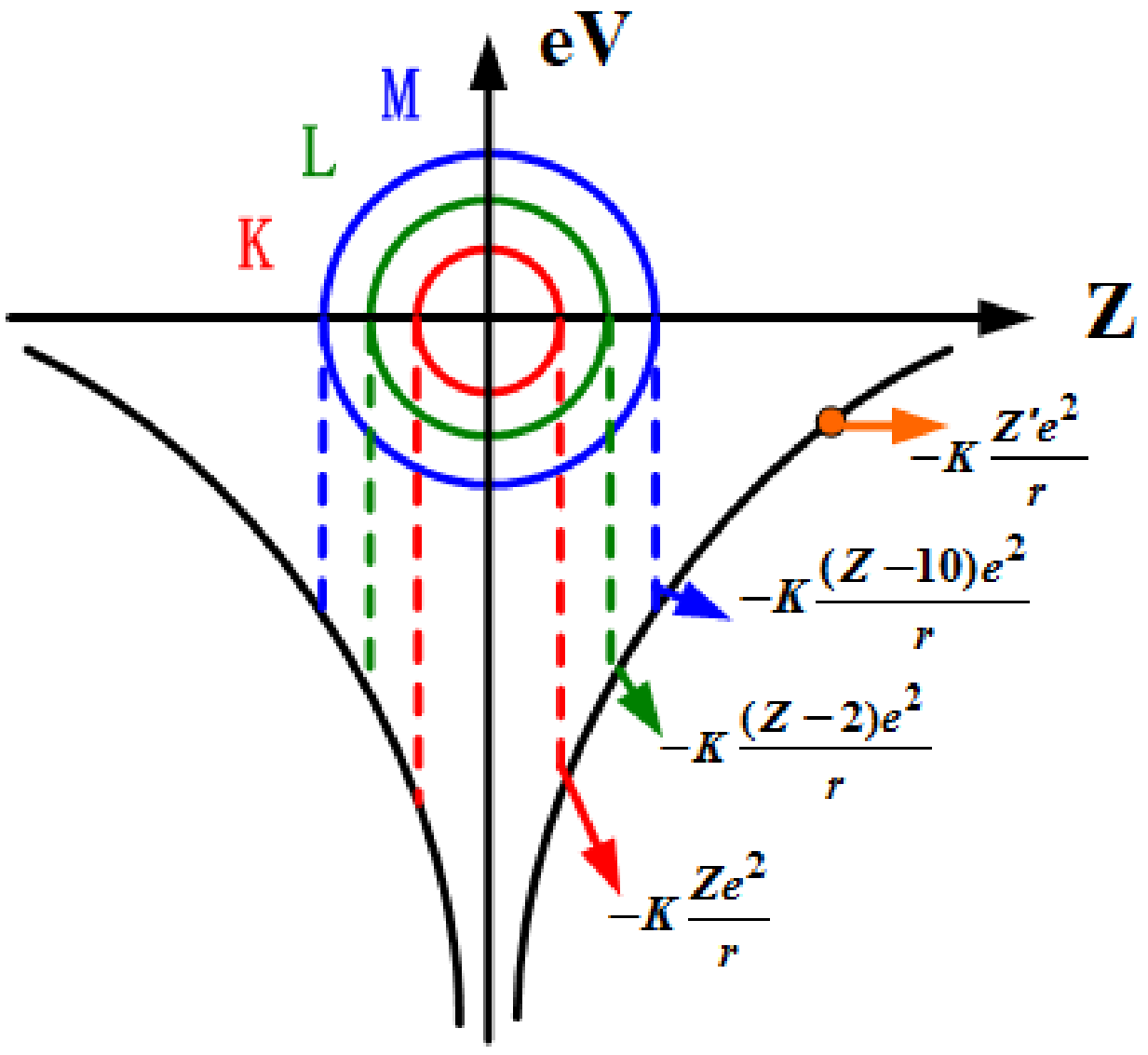

**Fig. 1** Schematic diagram of the atom potential energy, $K=\frac{1}{4\pi\varepsilon_0}$.

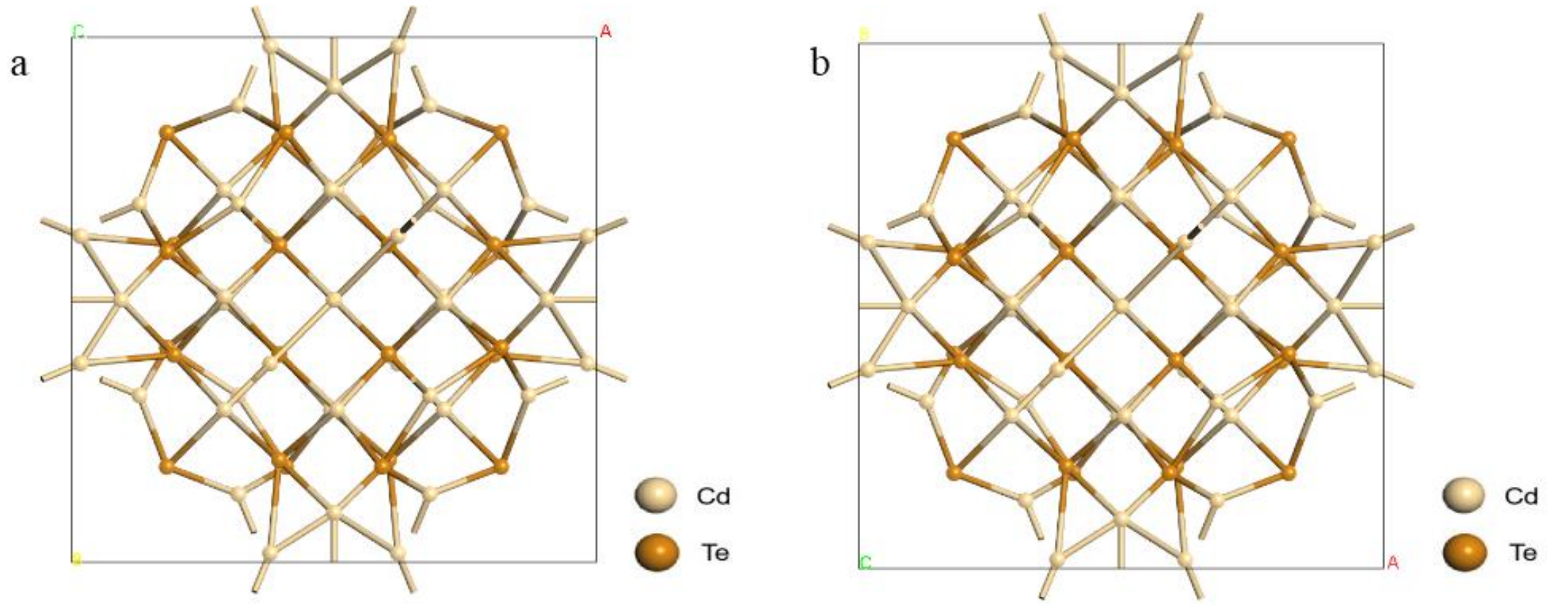

**Fig. 2** Optimized geometry of (a) Front view and (b) Top view $Cd_{43}Te_{28}$ of microporous materials.

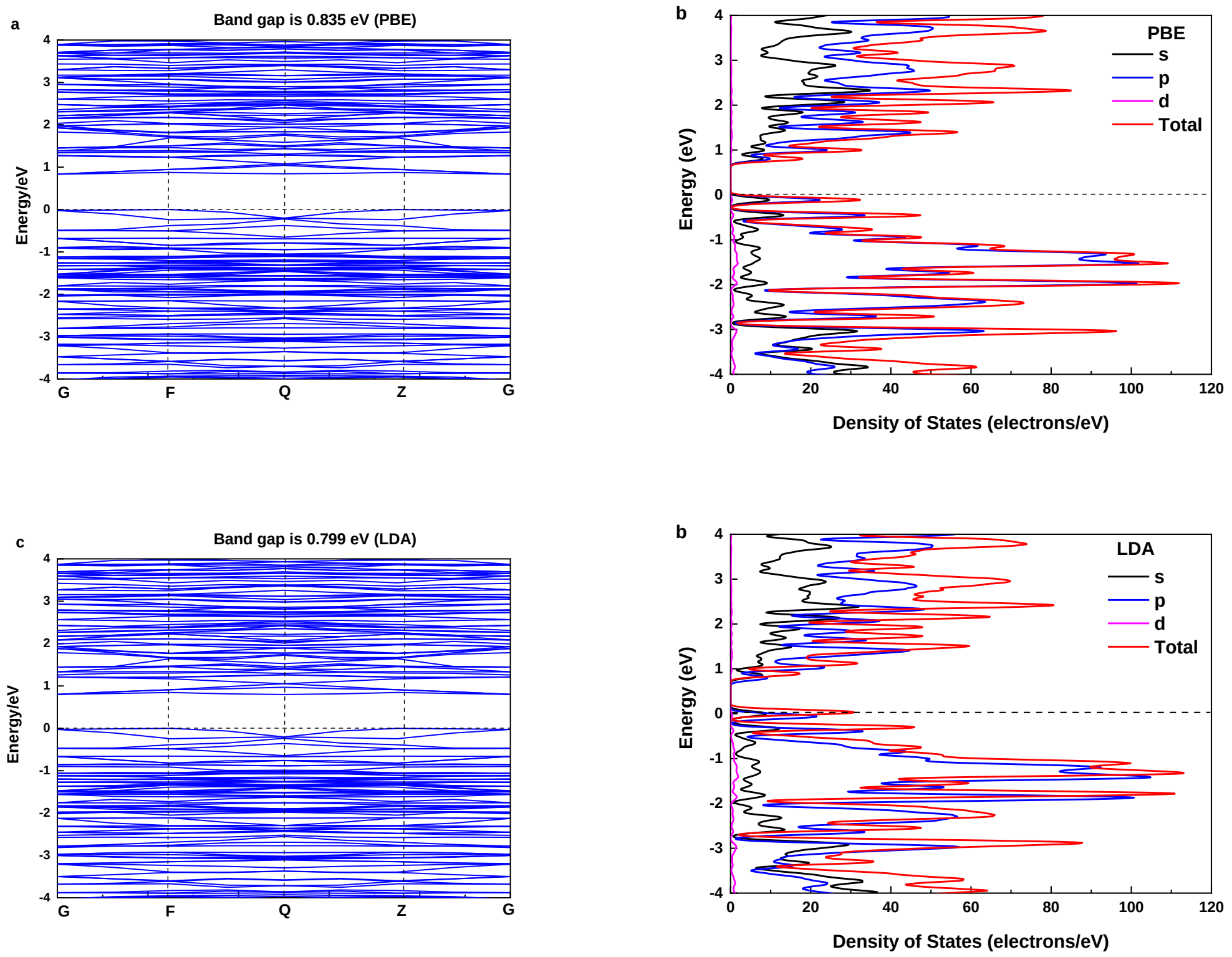

**Fig. 3** (a, c) Band structure and (b, d) partial DOS of $Cd_{43}Te_{28}$ microporous materials.

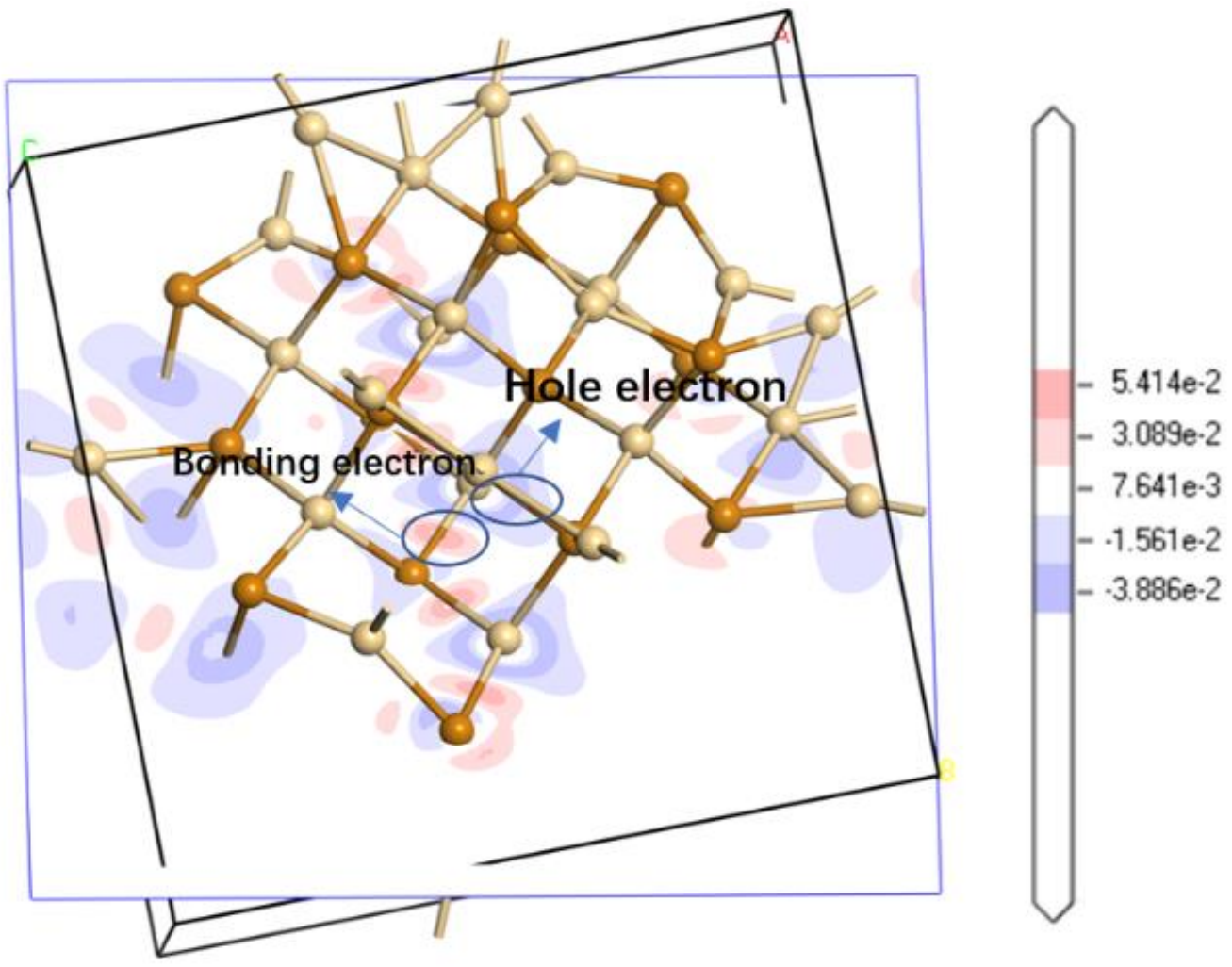

**Fig. 4** Deformation charge density of $Cd_{43}Te_{28}$ microporous materials.

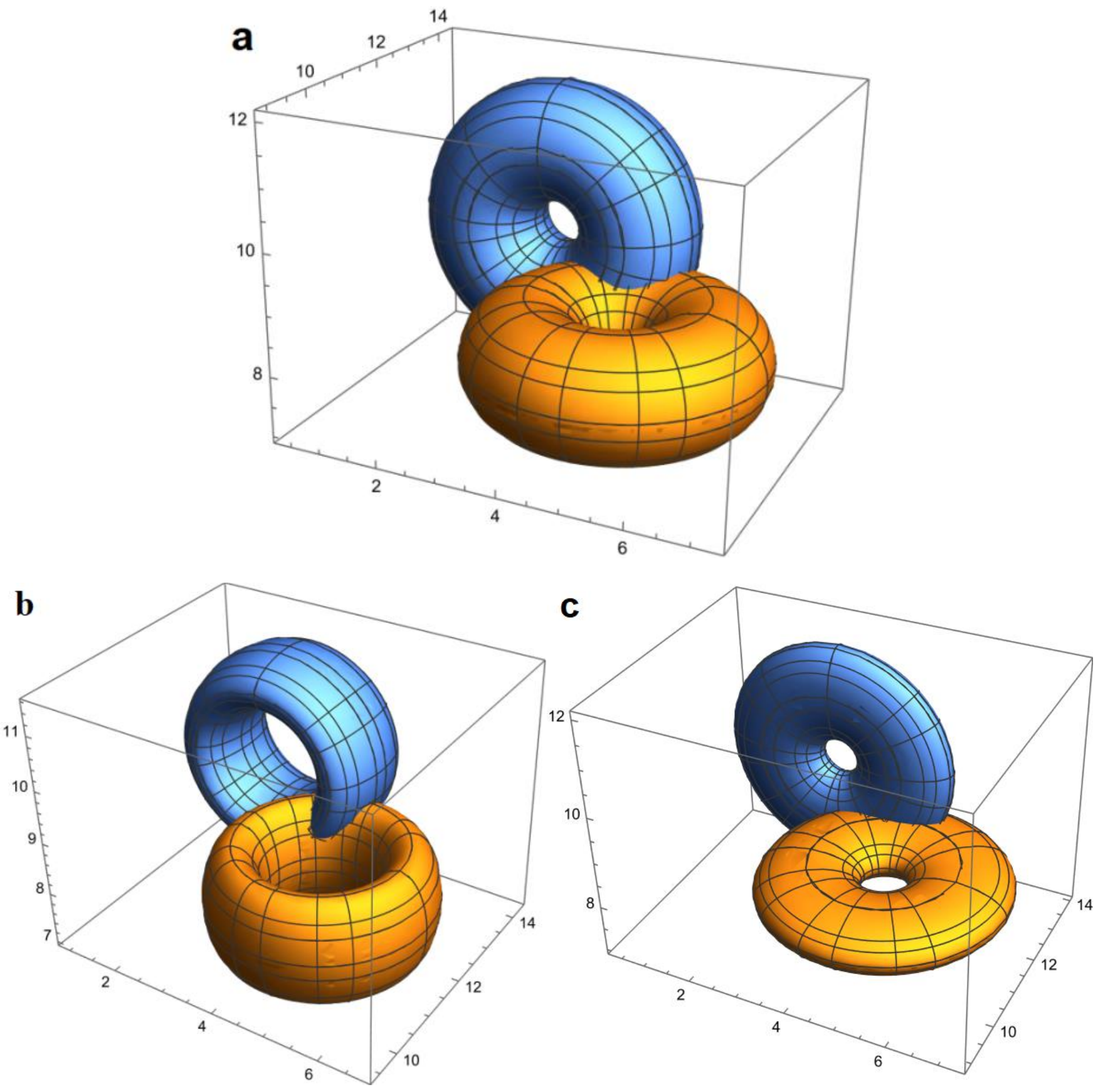

**Fig.5** The geometric topological bonding (a) A = B = 1, $\chi = 0$, $u$ = 6, $v$ = 6, (b) A = 0.5, B = 1, $\chi = 0$, $u$ = 6, $v$ = 6 and (c) A =1, B = 0.5, $\chi = 0$, $u$ = 6, $v$ = 6 of $Cd_{43}Te_{28}$ microporous materials using BBC model.

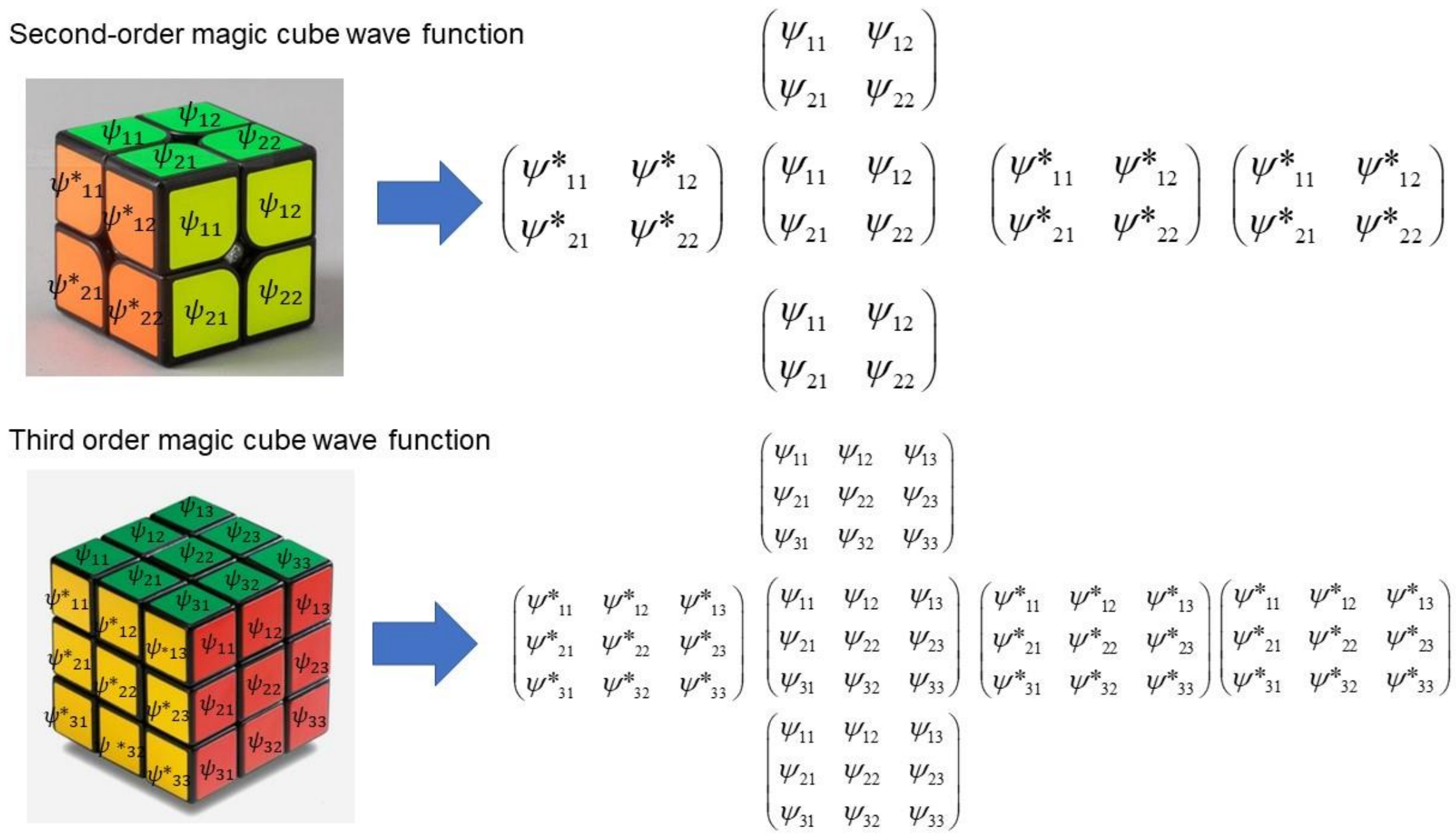

**Fig.6** Second- and Third- order magic cube wave function.

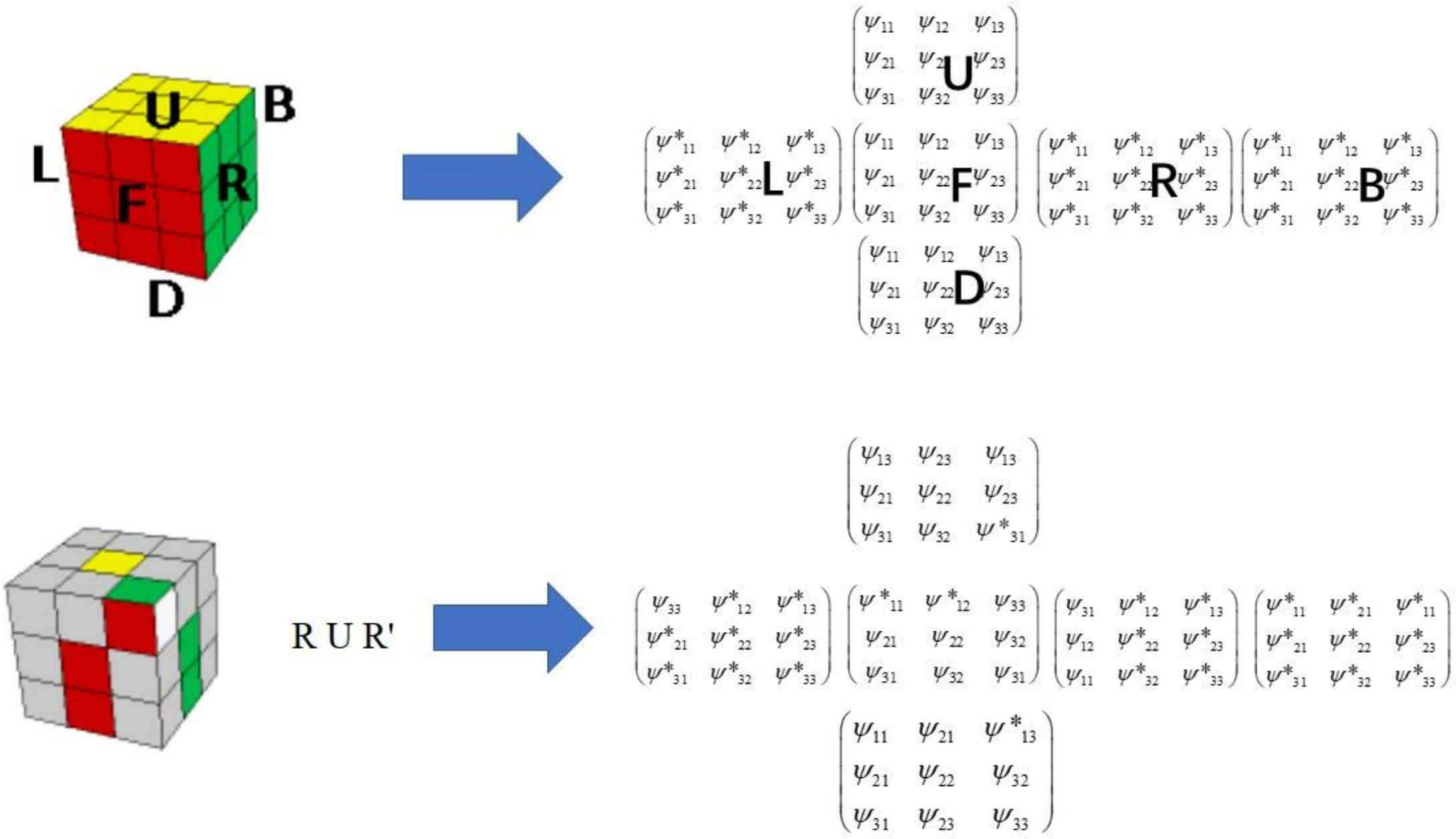

**Fig.7** The operation of the magic cube matrix wave function

**Table 1** Lattice parameters cutoff and k-point of $Cd_{43}Te_{28}$ microporous materials.

| Structure | Cutoff | k-point | Angles | | | Lattice parameters | | |
|---|---|---|---|---|---|---|---|---|
| | | | $\alpha$(deg) | $\beta$(deg) | $\gamma$(deg) | $a$(Å) | $b$(Å) | $c$(Å) |
| $Cd_{43}Te_{28}$ | 400 eV | 3*3*3 | 90.00° | 90.00° | 90.00° | 16.118 | 16.118 | 16.118 |

**Table 2** Laplace transform formula.

| Primitive function | Laplace transform | Convergence region |
|---|---|---|
| $f(t)=\zeta^{-1}\left[F(s)\right]$ | $F(s)=\zeta\left[f(t)\right]$ | |
| $\delta(t)$ | 1 | $\infty > s > -\infty$ |
| 1 | $\frac{1}{s}$ | $s > 0$ |
| $e^{-at}$ | $\frac{1}{s+a}$ | $s > -a$ |
| $\sin(at)$ | $\frac{a}{s^2+a^2}$ | $s > 0$ |
| $\cos(at)$ | $\frac{s}{s^2+a^2}$ | $s > 0$ |
| $\sinh(at)$ | $\frac{a}{s^2-a^2}$ | $s > \|a\|$ |
| $\cosh(at)$ | $\frac{s}{s^2-a^2}$ | $s > \|a\|$ |
| $e^{at}\sin(bt)$ | $\frac{b}{(s-a^2)+b^2}$ | $s > a$ |
| $e^{at}\cos(bt)$ | $\frac{s-a}{(s-a^2)+b^2}$ | $s > a$ |
| $e^{at}\sinh(bt)$ | $\frac{b}{(s-a^2)-b^2}$ | $s-a > \|b\|$ |
| $e^{at}\cos(bt)$ | $\frac{s-a}{(s-a^2)-b^2}$ | $s-a > \|b\|$ |

**Table 3** Atomic position coordinates obtained after optimization of $Cd_{43}Te_{28}$ microporous materials.

| $Cd_{43}Te_{28}$ | X | Y | Z |
|---|---|---|---|
| Cd1 | 15.8430 | 10.0174 | 10.0174 |
| Cd2 | 10.0174 | 15.8430 | 10.0174 |
| Cd3 | 10.0174 | 10.0174 | 15.8430 |
| Cd4 | 7.9649 | 11.4311 | 11.4311 |
| Cd5 | 5.1224 | 14.0294 | 10.9959 |
| Cd6 | 5.1224 | 10.9959 | 14.0294 |
| Cd7 | 2.0889 | 10.9959 | 10.9959 |
| Cd8 | 14.0294 | 5.1224 | 10.9959 |
| Cd9 | 11.4311 | 7.9649 | 11.4311 |
| Cd10 | 10.9959 | 5.1224 | 14.0294 |
| Cd11 | 8.0592 | 8.0592 | 14.6006 |
| Cd12 | 8.1534 | 4.6872 | 11.4311 |
| Cd13 | 4.6872 | 8.1534 | 11.4311 |
| Cd14 | 6.1009 | 6.1009 | 15.8430 |
| Cd15 | 0.2753 | 6.1009 | 10.0174 |
| Cd16 | 10.9959 | 2.0889 | 10.9959 |
| Cd17 | 6.1009 | 0.2753 | 10.0174 |
| Cd18 | 14.0294 | 10.9959 | 5.1224 |
| Cd19 | 10.9959 | 14.0294 | 5.1224 |
| Cd20 | 11.4311 | 11.4311 | 7.9649 |
| Cd21 | 8.0592 | 14.6006 | 8.0592 |
| Cd22 | 8.1534 | 11.4311 | 4.6872 |
| Cd23 | 6.1009 | 15.8430 | 6.1009 |
| Cd24 | 4.6872 | 11.4311 | 8.1534 |
| Cd25 | 0.2753 | 10.0174 | 6.1009 |
| Cd26 | 14.6006 | 8.0592 | 8.0592 |
| Cd27 | 15.8430 | 6.1009 | 6.1009 |
| Cd28 | 11.4311 | 8.1534 | 4.6872 |
| Cd29 | 11.4311 | 4.6872 | 8.1534 |
| Cd30 | 8.0592 | 8.0592 | 8.0592 |
| Cd31 | 7.9649 | 4.6872 | 4.6872 |
| Cd32 | 4.6872 | 7.9649 | 4.6872 |
| Cd33 | 4.6872 | 4.6872 | 7.9649 |
| Cd34 | 1.5177 | 8.0592 | 8.0592 |
| Cd35 | 2.0889 | 5.1224 | 5.1224 |
| Cd36 | 10.0174 | 0.2753 | 6.1009 |
| Cd37 | 8.0592 | 1.5177 | 8.0592 |
| Cd38 | 5.1224 | 2.0889 | 5.1224 |
| Cd39 | 10.9959 | 10.9959 | 2.0889 |
| Cd40 | 6.1009 | 10.0174 | 0.2753 |
| Cd41 | 10.0174 | 6.1009 | 0.2753 |
| Cd42 | 8.0592 | 8.0592 | 1.5177 |
| Cd43 | 5.1224 | 5.1224 | 2.0889 |
| Te1 | 12.9744 | 9.7538 | 9.7538 |
| Te2 | 9.7538 | 12.9744 | 9.7538 |
| Te3 | 9.7538 | 9.7538 | 12.9744 |
| Te4 | 6.5871 | 13.1914 | 13.1914 |
| Te5 | 6.3920 | 9.7263 | 9.7263 |
| Te6 | 2.9269 | 13.1914 | 9.5312 |
| Te7 | 2.9269 | 9.5312 | 13.1914 |
| Te8 | 13.1914 | 6.5871 | 13.1914 |
| Te9 | 13.1914 | 2.9269 | 9.5312 |
| Te10 | 9.7263 | 6.3920 | 9.7263 |
| Te11 | 9.5312 | 2.9269 | 13.1914 |
| Te12 | 6.3645 | 6.3645 | 12.9744 |
| Te13 | 6.3645 | 3.1439 | 9.7538 |
| Te14 | 3.1439 | 6.3645 | 9.7538 |
| Te15 | 13.1914 | 13.1914 | 6.5871 |
| Te16 | 13.1914 | 9.5312 | 2.9269 |
| Te17 | 9.5312 | 13.1914 | 2.9269 |
| Te18 | 9.7263 | 9.7263 | 6.3920 |
| Te19 | 6.3645 | 12.9744 | 6.3645 |
| Te20 | 6.3645 | 9.7538 | 3.1439 |
| Te21 | 3.1439 | 9.7538 | 6.3645 |
| Te22 | 12.9744 | 6.3645 | 6.3645 |
| Te23 | 9.7538 | 6.3645 | 3.1439 |
| Te24 | 9.7538 | 3.1439 | 6.3645 |
| Te25 | 6.3920 | 6.3920 | 6.3920 |
| Te26 | 6.5871 | 2.9269 | 2.9269 |
| Te27 | 2.9269 | 6.5871 | 2.9269 |
| Te28 | 2.9269 | 2.9269 | 6.5871 |

**Table 4** The deformation charge density $\delta\rho(\vec{r})$ and deformation charge of bond energy $\Delta V_{bc}(\vec{r}_{ij})$, as obtained by using BBC model.

$(\varepsilon_0 = 8.85\times10^{-12} C^2N^{-1}m^{-2}, e = 1.60\times10^{-19} C, r_i = 1.44\,\mathring{A}(\mathrm{Cd}), r_j = 1.38\,\mathring{A}(\mathrm{Te}), |\vec{r}_{ij}| = |\vec{r}_i - \vec{r}_j| = d_{ij}/2 = 1.49\,\mathring{A})$

| Structure | $\delta\rho^{hole-electron}(\vec{r}_i)\left(e/\mathring{A}^3\right)$ | $\delta\rho^{Bonding-electron}(\vec{r}_j)\left(e/\mathring{A}^3\right)$ | $\Delta V_{bc}^{bonding}(\vec{r}_{ij})(\mathrm{eV})$ |
|---|---|---|---|
| $Cd_{43}Te_{28}$ | -0.0389 | 0.0541 | -0.0797 |

**Table 5** The atomic position coordinates and the atomic radius *r* of $Cd_{43}Te_{28}$ microporous materials.

| | element | X | Y | Z | *r* |
|---|---|---|---|---|---|
| $Cd_{43}Te_{28}$ | Cd-24 | 4.6872 | 11.4311 | 8.1534 | 1.44 |
| | Te-6 | 2.9269 | 13.1914 | 9.5312 | 1.38 |

# Supplemental Material

# Topological Bonding and Electronic properties of $Cd_{43}Te_{28}$ semiconductor material with microporous structure

Yixin Li, Wei Xiong, Lei Li, Zhuoming Zhou, Chuang Yao, Zhongkai Huang, and Maolin Bo*

Key Laboratory of Extraordinary Bond Engineering and Advanced Materials Technology (EBEAM) of Chongqing, Yangtze Normal University, Chongqing 408100, China

*Corresponding Author: E-mail: bmlwd@yznu.edu.cn (Maolin Bo).

## 2.1 Supporting materials of structural stability

We used molecular dynamics (MD) to simulate the structural stability of $Cd_{43}Te_{28}$ microporous materials. MD simulations was performed for $Cd_{43}Te_{28}$ structures in the box containing 1971 atoms using periodic boundary conditions. The $Cd_{43}Te_{28}$ structure were obtained by performing NVT and NVE ensemble with time increments at 1 fs for 100 ps (the total iteration steps are 100,000) until the potential energy accomplished a stable value. The box size, volume, number of atoms of $Cd_{43}Te_{28}$ structure, as is shown in the **Table S1**. **Figs. S1**, show the geometries of $Cd_{43}Te_{28}$ structure at initial set temperatures of 298K, 500K, 700K and 900K, calculated using NVT ensemble. **Figs. S2**, show the geometries of $Cd_{43}Te_{28}$ microporous materials using NVE ensemble with initial setting temperatures at 298K, 500K, 700K and 900K.

**Table S1** Box size, volume and number of atoms of $Cd_{43}Te_{28}$ microporous materials.

| Structure | Number of atoms | Volume | Box size | | |
|---|---|---|---|---|---|
| | | | a(Å) | b(Å) | c(Å) |
| $Cd_{43}Te_{28}$ | 1917 | 113063.67 Å $^{3}$ | 48.35 | 48.35 | 48.35 |

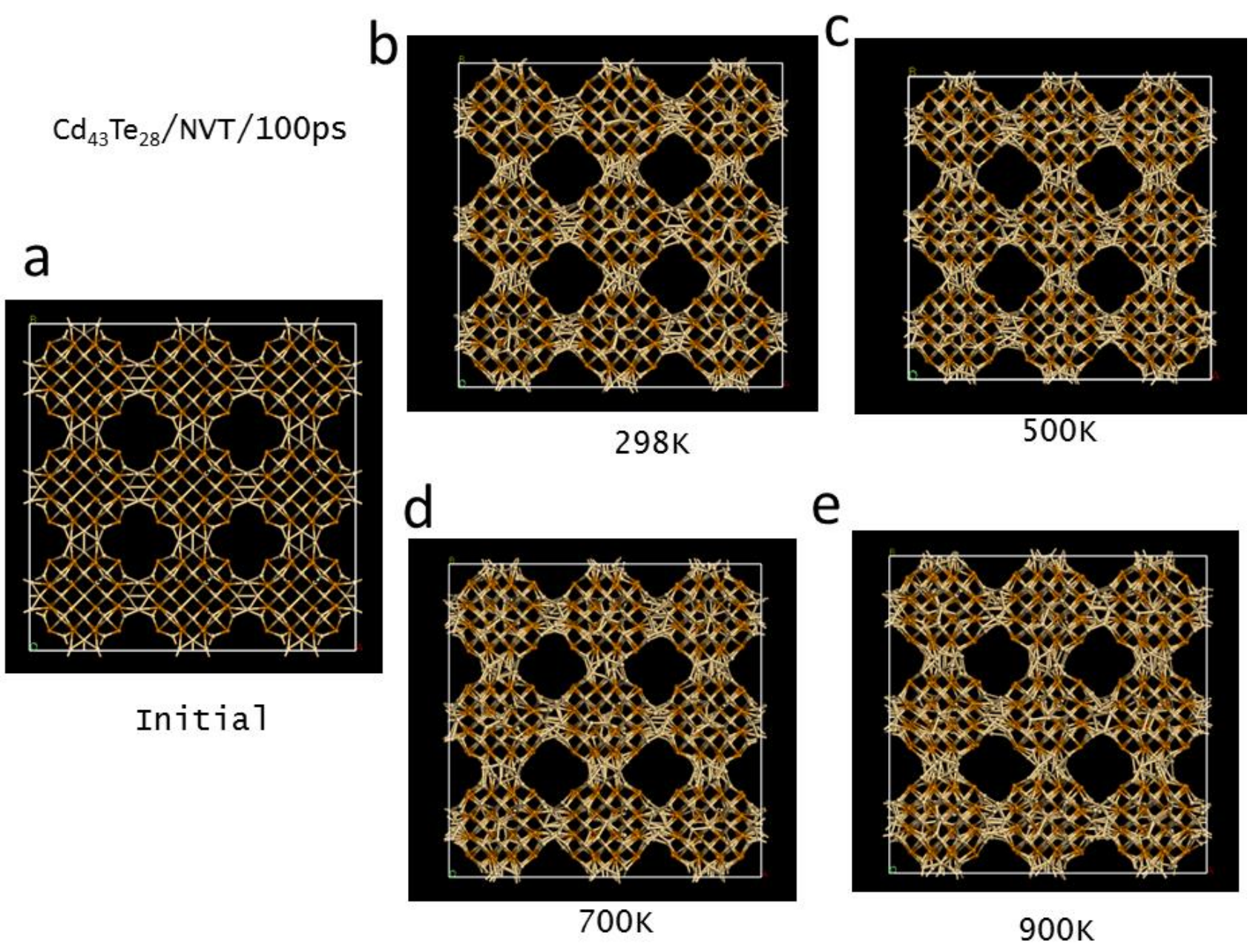

**Fig.S1** the geometries of $Cd_{43}Te_{28}$ semiconductor materials at initial set temperatures of 298K, 500K, 700K and 900K calculated using NVT ensemble.

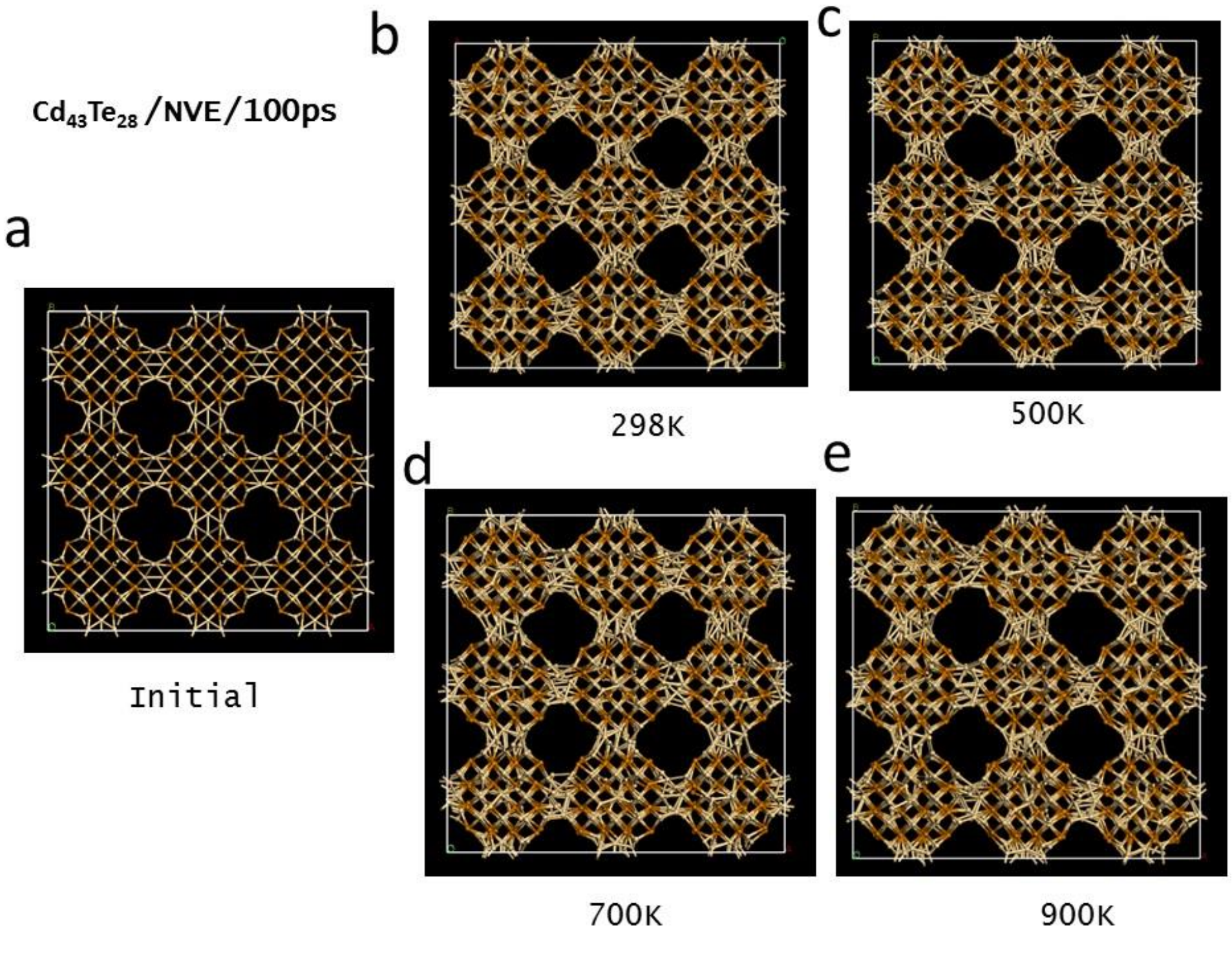

**Fig.S2** the geometries of $Cd_{43}Te_{28}$ semiconductor materials using NVE ensemble with initial setting temperatures at 298K, 500K, 700K and 900K.

## 2.2 Supporting materials of BBC model

### Supporting materials of BBC model

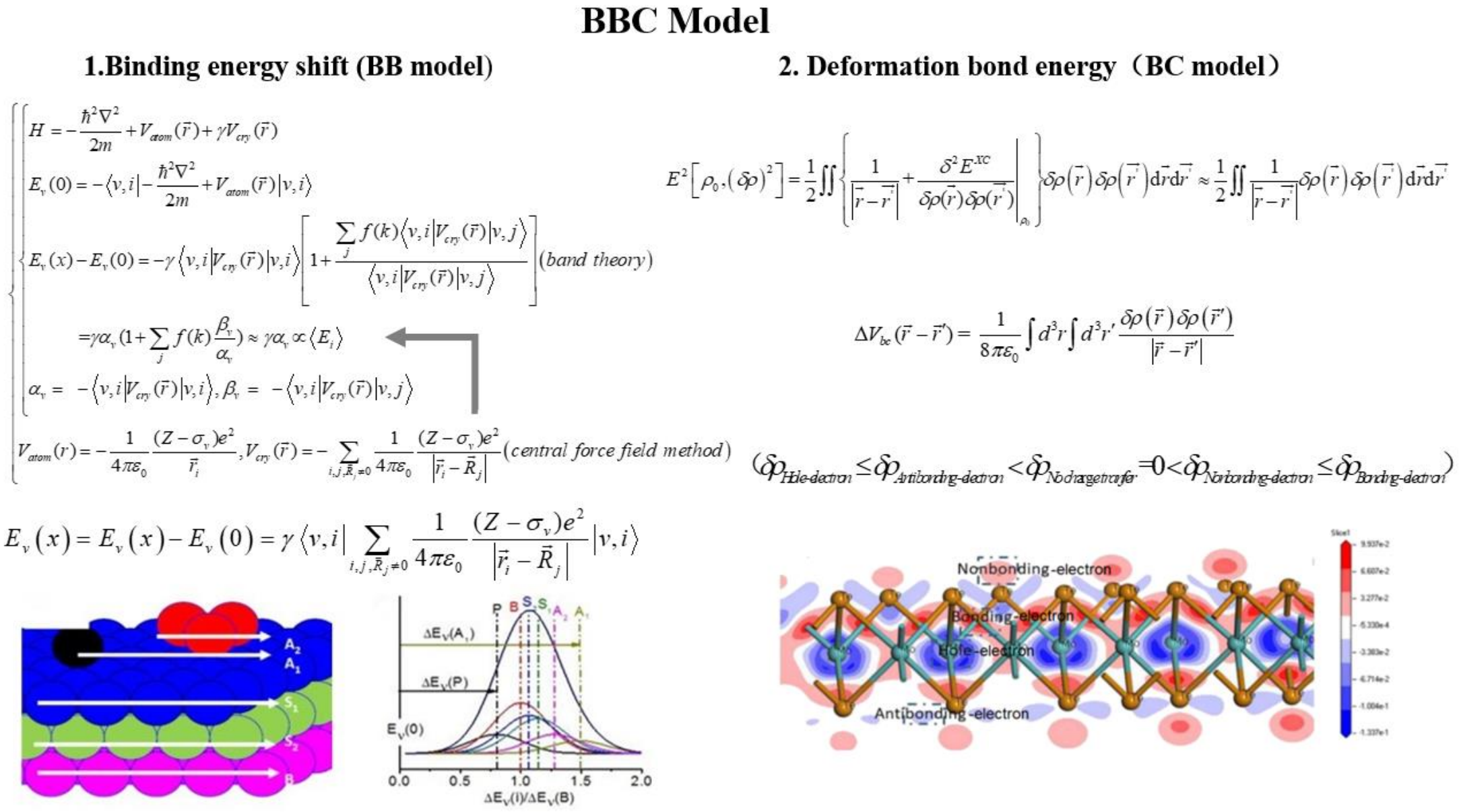

**Fig. S3** Schematic diagram of the BBC model.

The BBC model is quantified chemical bonds by binding energy shift and deformation charge density. For the binding energy (BB) model, we use the central field approximation and Tight-binding (TB) model, which can get the relationship between the energy shift and Hamiltonian. For the bond-charge (BC) model, we use the second-order term of energy expansion and use deformation charge density to calculate bonding states.

For binding energy (BB) model:

$$\begin{cases} H = \int \Psi^{+}(\vec{r})h(\vec{r})\Psi(\vec{r})\mathrm{d}r = \sum_{l,l'} C_l^{+}C_{l'}\int a^{*}(\vec{r}-\vec{l})h(\vec{r})a(\vec{r}-l')\mathrm{d}r \\ H = \xi(0)\sum_{l}\hat{n}_l - J\sum_{l}\sum_{\rho}C_l^{+}C_{l+\rho} \\ J = \sum_{l,l'} C_l^{+}C_{l'}\int a^{*}(\vec{r}-\vec{l})\left[V(\vec{r})-v_a(\vec{r}-\vec{l})\right]a(\vec{r}-\vec{l}')\mathrm{d}r \\ \sum_{l}\hat{n}_l = C_l^{+}C_l \end{cases} \tag{1}$$

$V(r)$ is the potential and $v_a(\vec{r}-\vec{l})$ is the atomic potential. $\xi(0)$ is the local

orbital electron energy. $\hat{n}_l$ represents the electron number operator in the Wanier representation.

$$\begin{cases} \xi(0)=\left(E_n^a - A_n\right) \\ A_n = -\int a_n^*\left(\vec{r}-\vec{l}\right)\left[V\left(\vec{r}\right)-v_a\left(\vec{r}-\vec{l}\right)\right]a_n\left(\vec{r}-\vec{l}\right)\mathrm{d}r \\ E_n^a=\int a_n^*\left(\vec{r}-\vec{l}\right)\left[-\frac{\hbar^2}{2m}\nabla^2+v_a\left(\vec{r}-\vec{l}\right)\right]a_n\left(\vec{r}-\vec{l}\right)\mathrm{d}r \\ \Psi(\vec{r})=\sum_l C_l a(\vec{r}-\vec{l}) \end{cases} \quad (2)$$

$E_n^a$ is the atomic energy level, and J is called the overlapping integral. $A_n$ represents the energy level shift, $E_n^a$ caused by the potential of (N-1) atoms outside the lattice point $l$ in the lattice

Consider the effect of external fields on energy level shifts:

$$\begin{cases} H'=\xi(0)\sum_l \hat{n}_l - J\sum_l\sum_\rho C_l^+ C_{l+\rho} = \left(E_n^a-\gamma A_n\right)\sum_l \hat{n}_l - \gamma J\sum_l\sum_\rho C_l^+ C_{l+\rho} \\ \Delta E_v(B)=A_n\sum_l \hat{n}_l, \ \Delta E_v(x)=\gamma A_n\sum_l \hat{n}_l \\ V_{cry}(\vec{r}_i-\vec{l}_j)=V\left(\vec{r}\right)-v_a\left(\vec{r}-\vec{l}\right) \\ \gamma V_{cry}(\vec{r}_i-\vec{l}_j)=\gamma\sum_{i,j,l_j\neq 0}\frac{1}{4\pi\varepsilon_0}\frac{Z'e^2}{\left|\vec{r}_i-\vec{l}_j\right|}=\gamma\sum_{i,j,l_j\neq 0}\frac{1}{4\pi\varepsilon_0}\frac{(Z-\sigma_v)e^2}{\left|\vec{r}_i-\vec{l}_j\right|} \\ \Delta E_v(x)=\gamma\int a_n^*\left(\vec{r}-\vec{l}\right)\sum_{i,j,l_j\neq 0}\frac{1}{4\pi\varepsilon_0}\frac{(Z-\sigma_v)e^2}{\left|\vec{r}_i-\vec{l}_j\right|}a_n\left(\vec{r}-\vec{l}\right)\mathrm{d}r\sum_l \hat{n}_l \\ \Delta E_v(x)=\gamma\left\langle\Psi_v\left(\vec{r}\right)\right|\sum_{i,j,\vec{R}_j\neq 0}\frac{1}{4\pi\varepsilon_0}\frac{(Z-\sigma_v)e^2}{\left|\vec{r}_i-\vec{l}_j\right|}\left|\Psi_v\left(\vec{r}\right)\right\rangle \end{cases} \quad (3)$$

The effective positive charge of the ion is $Z'=Z-\sigma_v$, considering the charge shielding effect $\sigma_v$, where $Z$ is the nuclear charge. $\Delta E_v(B)$ is represents the energy shift of an atom in an ideal bulk. $\delta\gamma=\gamma-1$ is relative bond energy ratio and $B$ indicates bulk atoms.

For bond-charge (BC) model， we consider the positive charge background ($b$) and the electron ($e$) as a system, and write their Hamiltonian sums and their interactions, respectively. In addition to electron kinetic energy, only electrostatic Coulomb interactions are considered:

$$
\begin{cases}
H = H_b + H_e + H_{eb} \\
H_e = \sum_{i=1}^{N} \frac{P_i^2}{2m} + \frac{1}{2} e_1^2 \sum_{i=1}^{N} \sum_{\substack{j=1 \\ i \neq j}}^{N} \frac{1}{\left|\vec{r}_i - \vec{r}_j\right|} e^{-\mu\left|\vec{r}_i - \vec{r}_j\right|} \\
H_b = \frac{1}{2} e_1^2 \int \mathrm{d}^3 x \int \mathrm{d}^3 x' \frac{n(\vec{x}) n(\vec{x}')}{\left|\vec{x} - \vec{x}'\right|} e^{-\mu\left|\vec{x} - \vec{x}'\right|} = \frac{1}{2} e_1^2 \left(\frac{N}{V}\right)^2 \int \mathrm{d}^3 x 4\pi \int \mathrm{d}z \frac{e^{-\mu z}}{z} = 4\pi e_1^2 \frac{N^2}{2V\mu^2} \\
H_{eb} = -e_1^2 \sum_{i=1}^{N} \int \mathrm{d}^3 x \frac{n(\vec{x})}{\left|\vec{x} - \vec{r}_i\right|} e^{-\mu\left|\vec{x} - \vec{r}\right|} = -e_1^2 \sum_{i=1}^{N} \frac{N}{V} 4\pi \int \mathrm{d}z \frac{e^{-\mu z}}{z} = -4\pi e_1^2 \frac{N^2}{V\mu^2}
\end{cases}
\tag{4}
$$

The shielding factor $\mu$ is added to the equation. $\vec{r}_i$ represents the *ith* electronic position. $\vec{x}$ represents the background position. $e$ is the basic charge, $e_1 = e/\sqrt{4\pi\varepsilon_0}$ . $e_1 n(\vec{x})$ is the charge density at background $\vec{x}$, and $n(\vec{x}) = N/V$ is a constant.

$$
\begin{cases}
H_e = \sum_{i=1}^{N} \frac{P_i^2}{2m} + \frac{1}{2} e_1^2 \sum_{i=1}^{N} \sum_{\substack{j=1 \\ i \neq j}}^{N} \frac{1}{\left|\vec{r}_i - \vec{r}_j\right|} e^{-\mu\left|\vec{r}_i - \vec{r}_j\right|} \\
\quad = \sum_{k\sigma} \frac{\hbar^2 k^2}{2m} a_{k\sigma}^{\dagger} a_{k\sigma} + \frac{e_1^2}{2V} \sum_{\vec{q}}{}^{*} \sum_{k\sigma} \sum_{k'\lambda} \frac{4\pi}{q^2 + \mu^2} a_{\vec{k}+\vec{q},\sigma}^{\dagger} a_{\vec{k}'-\vec{q},\lambda}^{\dagger} a_{\vec{k}'\lambda} a_{\vec{k}\sigma} + \frac{e_1^2}{2V} \frac{4\pi}{\mu^2} \left(N^2 - N\right) \\
\sum_{i=1}^{N} \frac{P^2}{2m} = \sum_{l'\sigma'} \sum_{l\sigma} a_{\vec{k}_{l'}\sigma'}^{\dagger} \left\langle \vec{k}_{l'}\sigma' \middle| \frac{P^2}{2m} \middle| \vec{k}_l \sigma \right\rangle a_{k_l\sigma} = \sum_{l\sigma} \frac{\hbar^2 \vec{k}_l^2}{2m} a_{\vec{k}_l\sigma}^{\dagger} a_{\vec{k}_l\sigma} \\
\frac{1}{2} e_1^2 \sum_{i=1}^{N} \sum_{\substack{j=1 \\ i \neq j}}^{N} \frac{1}{\left|\vec{r}_i - \vec{r}_j\right|} e^{-\mu\left|\vec{r}_i - \vec{r}_j\right|} = \frac{1}{2} e_1^2 \sum_{l'\sigma'} \sum_{m'\lambda'} \sum_{l\sigma} \sum_{m\lambda} a_{\vec{k}_{l'}\sigma'}^{?} a_{\vec{k}_{m'}\lambda'}^{\dagger} \left\langle \vec{k}_{l'}\sigma', \vec{k}_{m'}\lambda' \middle| \frac{e^{-\mu\left|\vec{r}_i - \vec{r}_j\right|}}{\left|\vec{r}_i - \vec{r}_j\right|} \middle| \vec{k}_l \sigma, \vec{k}_m \lambda \right\rangle a_{\vec{k}_m\lambda} a_{\vec{k}_l\sigma} \\
= \frac{1}{2} e_1^2 \sum_{l'} \sum_{m'} \sum_{l\sigma} \sum_{m\lambda} \delta_{\sigma\sigma'} \delta_{\lambda\lambda'} a_{\vec{k}_{l'}}^{?} a_{\vec{k}_{m'}}^{\dagger} \left\langle \vec{k}_{l'} \vec{k}_{m'} \middle| \frac{e^{-\mu\left|\vec{r}_i - \vec{r}_j\right|}}{\left|\vec{r}_i - \vec{r}_j\right|} \middle| \vec{k}_l \vec{k}_m \right\rangle a_{\vec{k}_m} a_{\vec{k}_l} \\
= \frac{e_1^2}{2V} \sum_{\vec{k}} \sum_{\vec{k}'} \sum_{\vec{q}} \sum_{\sigma\lambda} \frac{4\pi}{q^2 + \mu^2} a_{\vec{k}+\vec{q},\sigma}^{\dagger} a_{\vec{k}'-\vec{q},\lambda}^{\dagger} a_{\vec{k}'\lambda} a_{\vec{k}\sigma}
\end{cases}
\tag{5}
$$

In the formula, the $q$ on the sum sign of "$*$" indicates that the part where $q$ = 0 is ignored during the sum. When $V \to \infty, N \to \infty$, while keeping $N/V$ constant, the last term to the right of the equal sign causes the average energy $H_e/N$ of each particle to become:

$$\frac{1}{2}4\pi e_1^2\left(\frac{N}{V}\right)\frac{1}{\mu^2}-\frac{1}{2}4\pi e_1^2\left(\frac{N}{V}\right)\frac{1}{N}\frac{1}{\mu^2}$$

The former term is constant, and the latter term tends to zero. If $\mu\to 0$, the former term becomes a divergent term. However, this term just cancels out the divergent $H_b$ and $H_{eb}$ term. Thus, the Hamiltonian of the system becomes

$$H=\sum_{\vec{k}\sigma}\frac{\hbar^2\vec{k}^2}{2m}a_{k\sigma}^{\dagger}a_{k\sigma}+\frac{e_1^2}{2V}\sum_{\vec{q}}{}^{*}\sum_{\vec{k}\sigma}\sum_{\vec{k}'\lambda}\frac{4\pi}{q^2}a_{\vec{k}+\vec{q},\sigma}^{\dagger}a_{\vec{k}'-\vec{q},\lambda}^{\dagger}a_{\vec{k}'\lambda}a_{\vec{k}\sigma}$$

(6)

Electron interactions expressed using electron density:

$$\begin{cases}
\hat{V}_{ee}=\dfrac{e_1^2}{2V}\sum_{q}{}^{*}\sum_{\vec{k}\sigma}\sum_{\vec{k}'\lambda}\dfrac{4\pi}{q^2}a_{\vec{k}+\vec{q},\sigma}^{\dagger}a_{\vec{k}'-\vec{q},\lambda}^{\dagger}a_{\vec{k}'\lambda}a_{\vec{k}\sigma} \\
\quad=\dfrac{1}{2}e_1^2\sum_{l'\sigma'}\sum_{m'\lambda'}\sum_{l\sigma}\sum_{m\lambda}a_{\vec{k}_{l'}\sigma'}^{?}a_{\vec{k}_{m'}\lambda'}^{\dagger}\left\langle\vec{k}_{l'}\sigma',\vec{k}_{m'}\lambda'\left|\dfrac{1}{\left|\vec{r}_i-\vec{r}_j\right|}\right|\vec{k}_l\sigma,\vec{k}_m\lambda\right\rangle a_{\vec{k}_m\lambda}a_{\vec{k}_l\sigma} \\
\quad=\dfrac{1}{2}\sum_{\vec{k}_1\vec{k}_2,\vec{k}_1'\vec{k}_2'}\sum_{\sigma_1\sigma_2}\left\langle\vec{k}_1,\vec{k}_2\left|v\right|\vec{k}_1',\vec{k}_2'\right\rangle C_{\vec{k}_1\sigma_1}^{+}C_{\vec{k}_2\sigma_2}^{+}C_{\vec{k}_2'\sigma_2}C_{\vec{k}_1'\sigma_1}=\dfrac{1}{2}U\sum_{i}\sum_{\sigma\sigma'}C_{i\sigma}^{+}C_{i\sigma'}^{+}C_{i\sigma'}C_{i\sigma}==\dfrac{1}{2}U\sum_{i}\sum_{\sigma\sigma'}n_{i\sigma}n_{i\bar{\sigma}} \\
\quad=\dfrac{1}{4\pi\varepsilon_0}\times\dfrac{1}{2\left|\vec{r}-\vec{r}'\right|}\int d^3r\int d^3r'\rho\left(\vec{r}\right)\rho\left(\vec{r}'\right) \\
\rho\left(\vec{r}\right)=\alpha_{\zeta}^{+}\left(\vec{r}\right)\alpha_{\zeta}^{-}\left(\vec{r}\right),\alpha_{\zeta}^{+}\left(\vec{r}\right)=\sum_{\vec{k}}\psi_{\vec{k}}^{*}\left(\vec{r}\right)a_{\vec{k}\zeta}^{+}\equiv\sum_{i}\psi_{\vec{k}i}^{*}\left(\vec{r}\right)a_{i\zeta}^{+} \\
\left\langle k_1,k_2\left|v\right|k_1',k_2'\right\rangle=\int\dfrac{e^2\psi_{k_1}^{*}\left(\vec{r}\right)\psi_{k_2}^{*}\left(\vec{r}'\right)\psi_{k_1'}\left(\vec{r}\right)\psi_{k_2'}\left(\vec{r}'\right)}{4\pi\varepsilon_0\left|\vec{r}-\vec{r}'\right|}\mathrm{d}r\mathrm{d}r'
\end{cases}$$

(7)

Electron interaction terms for density fluctuations:

$$
\begin{cases}
\delta V_{ee} = \dfrac{1}{4\pi\varepsilon_0} \times \dfrac{1}{2\left|\vec{r} - \vec{r}'\right|} \int d^3 r \int d^3 r' \delta\rho\left(\vec{r}\right) \delta\rho\left(\vec{r}'\right) \\
\quad = \pm \dfrac{1}{4\pi\varepsilon_0} \times \dfrac{1}{2\left|\vec{r} - \vec{r}'\right|} \int d^3 r \int d^3 r' \rho\left(\vec{r}\right) \rho\left(\vec{r}'\right) e^{-\mu(\vec{r}-\vec{r}')} \\
\quad = \pm \dfrac{e_1^2}{2V} \sum_{\vec{k}} \sum_{\vec{k}'} \sum_{\vec{q}} \sum_{\sigma\lambda} \dfrac{4\pi}{q^2 + \mu^2} a^{\dagger}_{\vec{k}+\vec{q},\sigma} a^{\dagger}_{\vec{k}'-\vec{q},\lambda} a_{\vec{k}'\lambda} a_{\vec{k}\sigma} \\
\Delta V_{bc}(\vec{r} - \vec{r}') = \delta V_{ee} = \dfrac{1}{4\pi\varepsilon_0} \dfrac{1}{2\left|\vec{r} - \vec{r}'\right|} \int d^3 r \int d^3 r' \delta\rho\left(\vec{r}\right) \delta\rho\left(\vec{r}'\right)
\end{cases}
\tag{8}
$$